\documentclass[]{aa}
\usepackage{graphicx}
\usepackage{txfonts}
\usepackage{natbib}

\bibpunct{(}{)}{;}{a}{}{,} 
\begin{document} 

   \title{{\it Herschel}\thanks{{\it Herschel} is an ESA space observatory with science instruments provided by European--led Principal Investigator consortia and with important participation from NASA.} Water Maps towards the vicinity of the black hole Sgr A*}


   \author{J. Armijos-Abenda\~no\inst{1,2}
          \and
          J. Mart\'in-Pintado\inst{2}
          \and
          M. A. Requena-Torres\inst{3,4}
          \and
          E. Gonz\'alez-Alfonso\inst{5}
          \and
          R. G\"usten\inst{4} 
          \and 
          A. Wei\ss\inst{4} 
          \and 
          A. I. Harris\inst{3} 
          \and 
          F. P. Israel\inst{6} 
          \and 
          C. Kramer\inst{7}
          \and 
          J. Stutzki\inst{8} 
          \and
          P. van der Werf\inst{6}
          }
   \institute{Observatorio Astronómico de Quito, Escuela Polit\'ecnica Nacional, Av. Gran Colombia S/N, Interior del Parque La Alameda, 170136, Quito, Ecuador
         \and
             Centro de Astrobiolog\'ia (CSIC, INTA), Ctra a Ajalvir, km 4, 28850, Torrej\'on de Ardoz, Madrid, Spain
         \and
            Department of Astronomy, University of Maryland, College Park, MD 20742, USA
         \and 
             Max-Planck Institut f\"ur Radioastronomie, Auf dem H\"ugel 69, D-53121 Bonn, Germany
         \and
             Universidad de Alcal\'a de Henares, Departamento de F\'isica, Campus Universitario, E-28871 Alcal\'a de Henares, Madrid, Spain
         \and
             Leiden Observatory, Leiden University, P.O. Box 9513, 2300 RA Leiden, The Netherlands
         \and
             IRAM, Avenida Divina Pastora 7, 18012 Granada, Spain
         \and
            KOSMA, I. Phsikalisches Institut der Universit\"at zu K\"oln, Z\"ulpicher Strasse 77, 50937 K\"oln, Germany}
            
   \date{Received September 01, 2017; accepted January 15, 2018}

 
  \abstract
    {}
   {We study the spatial distribution and kinematics of water emission in a $\sim$8$\times$8 pc$^2$ region of the Galactic Center (GC) that covers the main molecular 
   features around the supermassive black hole Sagittarius A$^*$ (Sgr A$^*$). We also analyze the water excitation to derive the physical conditions 
   and water abundances in the Circumnuclear Disk (CND) and the ``quiescent clouds''.}
   {We presented the integrated line intensity maps of the ortho 1$_{10}-1_{01}$, and para 2$_{02}-1_{11}$ and 1$_{11}-0_{00}$ water transitions observed using the OTF mapping mode with the HIFI instrument on board 
   {\it Herschel}. To study the water excitation we used HIFI observations of the ground state ortho and para H$_2^{18}$O transitions towards three 
   selected positions in the vicinity of Sgr A$^*$. In our study, we also used dust continuum measurements of the CND, obtained with the SPIRE instrument.
   Using a non--LTE radiative transfer code, the water line profiles and dust continuum were modeled, deriving
   H$_2$O abundances (X$_{\rm H_2O}$), turbulent velocities (V$_{\rm t}$) and dust temperatures (T$_{\rm d}$).
   We also used a rotating ring model to reproduce the CND kinematics represented by the Position Velocity (PV) diagram derived from para 2$_{02}-1_{11}$ H$_2$O lines.}
   {In our H$_2$O maps we identify the emission associated with known features around Sgr A*: CND, the Western Streamer, and the 20 and 50 km s$^{-1}$ clouds.
    The ground-state ortho water maps show absorption structures in the velocity range of [-220,10] km s$^{-1}$ associated with foreground sources.
    The PV diagram reveals that the 2$_{02}-1_{11}$ H$_2$O emission traces the CND also observed in 
    other high--dipole molecules such as SiO, HCN and CN. Using the non--LTE code, we derive high X$_{\rm H_2O}$ of $\sim$(0.1--1.3)$\times$10$^{-5}$, V$_{\rm t}$ of 14--23 km s$^{-1}$ and T$_{\rm d}$ of 15--45 K for the CND, and the lower X$_{\rm H_2O}$ of 4$\times$10$^{-8}$ and V$_{\rm t}$ of 9 km s$^{-1}$ for the 20 km s$^{-1}$ cloud. Collisional excitation and dust effects are responsible for the water excitation in the southwest lobe of the CND and the 20 km s$^{-1}$ cloud, whereas only collisions can account for the water excitation in the northeast lobe of the CND. We propose that the water vapor in the CND is produced by grain sputtering by shocks of 10--20 km s$^{-1}$, with some contribution of high temperature and cosmic--ray chemistries plus a PDR chemistry, whereas the low X$_{\rm H_2O}$ derived for the 20 km s$^{-1}$ cloud could be partially a consequence of the water freeze--out on grains.}
   {}


   \maketitle
%

\section{Introduction}\label{Intro}
\subsection{Main features within 8 pc around Sgr A$^*$}
\begin{figure}
\centering
\includegraphics[width=7cm]{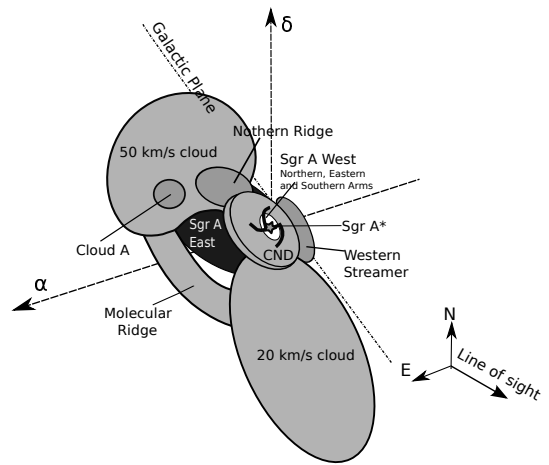}
\caption{Sketch of the main features within the 200$\times$200 arcsec$^2$ region around the suppermassive black hole Sgr A$^*$ shown by a star. 
The features indicated in gray correspond to mostly the molecular components. The Sgr A East SNR (black ellipse) and Sgr A West (black minispiral) are 
shown as well. The ionized Northern, Eastern and Southern arms of Sgr A West are indicated. The big circle and ellipse represent 
the 50 and 20 km s$^{-1}$ clouds, respectively. Both clouds seem to be connected by the Molecular Ridge (the curved streamer). The Western Streamer 
and the Northern Ridge are also shown.}
\label{fig1}
\end{figure}

The Galactic Center (GC) has been the subject of many multifrequency studies due to the large variety of processes taking place in this special region of 
the Galaxy. The GC interstellar medium (ISM) is affected by high energy phenomena \citep{Koyama,Wang,Terrier,Ponti}, large scale shocks \citep{Pinta01} 
and star formation \citep{Gaume,DePree,Blum,Paumard}. The main GC molecular clouds within the $\sim$200$\times$200 arcsec$^2$ region ($\sim$8$\times$8
pc$^2$ at a GC distance of 7.9 kpc \citep{Boehle}) around the supermassive black hole Sgr A$^*$ are sketched in Fig.~\ref{fig1}. 
Sgr A$^*$ is surrounded by Sgr A West, that is composed of three ionized 
filaments: the Northern, Eastern and Southern Arms \citep{Yusef93}. These filaments could be streamers of ionized gas feeding Sgr A$^*$ \citep{Zhao}.

Sgr A West is surrounded by an inclined and clumpy Circumnuclear Disk (CND) of gas and dust, which has inner and outer edges around 2 and 5 pc, respectively. It has an inclination around 63$^{\circ}$ and rotates with a constant velocity of 110 km s$^{-1}$ \citep{Gusten1}. 
Using the CO(7--6) emission, \citet{Harris} derived an H$_2$ density of $\sim$3$\times$10$^4$ cm$^{-3}$ and a temperature of $\sim$300 K for the CND.
For this source, \citet{Oka} derived H$_2$ masses of $\approx$(2.3--5.2)$\times$10$^5$ and $\approx$5.7$\times$10$^6$ M$_{\sun}$
based on $^{13}$CO(1--0) intensity measurements and virial assumptions, respectively. Considering
the discrepancy in the H$_2$ mass estimates found by \citet{Oka}, \citet{Ferriere} disregarded the virial H$_2$ mass estimates for the CND, and derived an
H$_2$ mass of 2$\times$10$^5$ M$_{\sun}$ for this source based on measurements of $^{12}$CO and $^{13}$CO ground--level transitions. 
The CO excitation in the CND has been studied by \citet{Reque12} utilizing a large velocity gradient (LVG) model. 
They derived temperatures of $\sim$200 K and H$_2$ densities of 
$\sim$3.2$\times$10$^4$ cm$^{-3}$ for the bulk of the CND material, confirming its transient nature. This was also confirmed with dense gas tracers as 
HCN and HCO$^+$ using the APEX telescope \citep{Betsy}.

The inner central parsec around the black hole known as the central cavity has been characterized by a hot CO gas component with a temperature around 10$^{3.1}$ K and a H$_{\rm 2}$ density $\lesssim$10$^4$ cm$^{-3}$ or with multiple cooler components ($\lesssim$300 K) at higher densities \citep{Goicoe2013}. UV radiation and shocks could heat this molecular gas if there is a small filling factor of clumps/clouds \citep{Goicoe2013}. A recent study showed the presence of a high positive--velocity gas in the central cavity with temperatures from 400 K to 2000 K and H$_2$ densities of (0.2-1.0)$\times$10$^5$ cm$^{-3}$ \citep{Goicoe2018b}.

A few parsecs from Sgr A$^*$ there are two giant molecular clouds, the 20 and 50 km s$^{-1}$ clouds.
\cite{Zylka} characterized the 20 km s$^{-1}$ cloud as a $\sim$15 pc$\times$(7.5 pc)$^2$ ellipsoid and the 50 km s$^{-1}$ cloud as having a size around 15 pc. These two clouds seem to be connected by a ridge of gas and dust, the Molecular Ridge \citep{Ho}. \citet{Maeda}
proposed however that the Molecular Ridge is part of the 50 km s$^{-1}$ cloud that has been compressed by the forward shock of an expanding shell of 
synchrotron emission, the Sgr A East supernova remnant (SNR).

It has been proposed that Sgr A East is located behind Sgr A$^*$ and the CND \citep{Coil}, while the 20 km s$^{-1}$ cloud lies in front of Sgr A$^*$, the CND and Sgr A East \citep{Herrnstein1,Park,Coil}. It is also thought that part of the 50 km s$^{-1}$ cloud lies behind Sgr A East \citep{Ferriere}, and a long and filamentary structure of gas and dust known as the Western Streamer borders the western edge of Sgr A East. Based on NH$_3$ images, \citet{Herrnstein} proposed that the expanding shell of Sgr A East is impacting the 50 km s$^{-1}$ cloud in the west and the Western Streamer in the east.

The Northern Ridge is a cloud that lies along the northern boundary of Sgr A East \citep{Ferriere}. 
Using NH$_3$ images, \citet{McGary} suggested that many filamentary features like the Northern Ridge are connecting the CND with the 50 km s$^{-1}$ cloud, indicating
that the clouds are most likely feeding the nucleus of the Galaxy. However, \citet{Ferriere} proposed that the Western Streamer and the Northern Ridge could be made of material swept--up by the expansion of Sgr A East.

Southwest from the 50 km s$^{-1}$ cloud, \citet{Baladron} observed SiO(2--1) emission of an isolated molecular cloud called Cloud A. They also found high 
HNCO abundances in the 20 and 50 km s$^{-1}$ clouds and the lowest HNCO abundances in the CND, whereas SiO showed high abundances towards both 
clouds and the CND. \citet{Baladron} proposed that the HNCO in the CND is being photodissociated by UV radiation from the central parsec star cluster \citep{Morris}. 
In the CND the SiO seems to be more 
resistant against UV--photons and/or is being produced very efficiently by the destruction of the grain cores due to strong shocks \citep{Baladron}.

\subsection{Water in the Galactic Center}
Water emission has been observed with high angular resolution towards the GC mainly through its maser emission. The 22 GHz line is almost exclusively seen as a maser. \citet{Yusef95} found four masers of 6$_{16}$-5$_{23}$ H$_2$O at 22 GHz within the inner 12 pc of the GC, one of them likely associated with a high-mass star-forming region and located at the boundary between Sgr A East and the 50 km s$^{-1}$ cloud. Using VLA observations \citet{Sjouwerm} detected eight 22 GHz H$_2$O masers in the 20 km s$^{-1}$ cloud. Also, 22 GHz H$_2$O masers have been found in the CND \citep{Yusef08}.

At low spatial resolution, using SWAS, \citet{Neufeld} observed widespread emission and absorption of the ortho 1$_{10}-1_{01}$ H$_2$O transition at 556.936 GHz in the strong submillimeter continuum source Sgr B2.
Furthermore, utilizing the Odin satellite \citet{Sandqvist} observed the ortho 1$_{10}$-1$_{01}$ H$_2$O emission towards the CND and the 20 and 50 km s$^{-1}$ clouds. They found ortho 1$_{10}-1_{01}$ H$_2$O absorption features at negative velocities associated with the 
Local Sgr, -30 km s$^{-1}$, the 3--kpc Galactic arms and the near side of the Molecular Ring surrounding the GC.
These absorption features are also detected in the 1$_{11}-0_{00}$ H$_2$O spectra observed towards the 20 and 50 km s$^{-1}$ clouds by \cite{Sonner13}. The water abundance of 5$\times$10$^{-8}$ is derived for foreground clouds located in the 3--kpc Galactic arm, while water abundances higher than 1.5$\times$10$^{-7}$ are measured for gas components with velocities $\leqslant$-85 km s$^{-1}$ located within the 200 pc region of Sgr A. Shocks or turbulent dissipation are proposed as the most likely mechanisms responsible for the origin of water in the Sgr A gas components with velocities $\leqslant$-85 km s$^{-1}$ \citep{Sonner13}.

Rotational excited and ground--state absorption lines of water are detected towards the central cavity, containing a hot molecular component heated by UV photons and shocks if there is a small filling factor of dense clumps/clouds \citep{Goicoe2013}. The central cavity shows a high positive--velocity wing in the 1$_{10}$-1$_{01}$ H$_2$O line, reaching velocities up to +270 km s$^{-1}$ \citep{Goicoe2018b}. The water in this region is thought to be originated in gas with elevated temperatures via gas--phase routes.

So far the H$_2$O emission/absorption distribution around Sgr A$^*$ has not been studied.
The GC offers an unique opportunity to study the excitation and origin of water in gas phase in typical GC clouds surrounding a supermassive black hole. As part of the {\it Herschel} EXtraGALactic (HEXGAL) guaranteed time program we mapped an area of $\sim$47$\times$47 pc$^2$ around Sgr A$^*$ in four H$_2$O lines (557, 988, 1113 and 1670 GHz) in order
to study the spatial distribution of the water and its kinematics in the vicinity of Sgr A$^*$. In this paper we have focused in the study of only an area of 
$\sim$8$\times$8 pc$^2$ around Sgr A$^*$. Furthermore, single position observations of ortho 1$_{10}-1_{01}$ and para 1$_{11}-0_{00}$ H$_2^{18}$O
transitions were observed as well, with the aim of better constraining the column density of water in this very complex region. In Sec. \ref{Obser} we present our observations. Maps and spectra of H$_2$O are presented in Sec. \ref{Results1}.
In Sec. \ref{Kinematic}, we study the kinematics of water in the surroundings of Sgr A$^*$.
The modeling of water and dust continuum emission using a non--LTE radiative transfer code and our results are described in Sec. \ref{Model}.
In Sec. \ref{discu5}, we discuss the excitation and chemistry/heating of water. Finally, we present the conclusions in Sec. \ref{Conclusions}.


\section{Observations}\label{Obser}
\begin{table}
\centering
\caption{Source positions}
\begin{tabular}{ccc}
\hline
\hline
Position & $\alpha$(J2000) & $\delta$(J2000)\\
\hline
CND$_1$     & 17$^{\rm h}$45$'$38$''$.43 & -29$^{\circ}$00\arcmin58\arcsec.10\\
CND$_2$     & 17$^{\rm h}$45$'$41$''$.85 & -28$^{\circ}$59\arcmin48\arcsec.20\\
20 km s$^{-1}$ cloud & 17$^{\rm h}$45$'$39$''$.95 & -29$^{\circ}$03\arcmin10\arcsec.00\\
\hline
\end{tabular}
\label{table01}
\end{table}

\begin{table*}
\centering
\caption{H$_2$O and H$_2^{18}$O observations}
\begin{tabular}{cccccc}
\hline
\hline
Species & Frequency & Observation & Observation & HIFI & HPBW   \\
  & (GHz)  &  date  & IDs & band & (\arcsec) \\
\hline
ortho--H$_2^{18}$O & 548 & September 2010 & 1342205521, 1342205522, 1342205523 & 1 &39\\
ortho--H$_2$O      & 557 & October 2010 & 1342205305, 1342206366, 1342206367 & 1 &38\\
para--H$_2$O      & 988 & March 2011 & 1342216817, 1342216818, 1342216819 & 4 &22\\
para--H$_2$O      &1113 & October 2010 & 1342206392, 1342206393, 1342206394 & 4 &19\\
para--H$_2^{18}$O &1102 & October 2010 & 1342206389, 1342206390, 1342206391 & 4 &19\\
ortho--H$_2$O      &1670 & February 2011 & 1342214462, 1342214463, 1342214464 & 6 &13\\
\hline
\end{tabular}
\label{table001}
\end{table*}

\begin{figure}
\centering
\includegraphics[width=9cm]{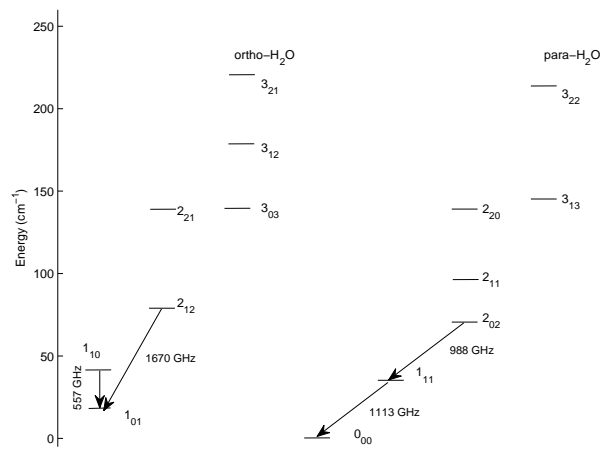}
\caption{Energy level diagram of ortho and para H$_2$O. The observed ortho and para H$_2$O transitions are shown with arrows. The 
frequencies of the observed H$_2$O transitions also are indicated.}
\label{fig2}
\end{figure}

The data were taken with the HIFI instrument \citep{De Graauw} on board the {\it Herschel} Space Observatory. 
Fig.~\ref{fig2} shows the energy level diagram of the observed H$_2$O transitions.
We performed mapping observations of the ortho 556.936 GHz (557 GHz) 1$_{10}-1_{01}$, para 987.927 GHz (988 GHz) 2$_{02}-1_{11}$, and para 1113.343 GHz (1113 GHz) 
1$_{11}-0_{00}$ transitions of H$_2$O. We mapped an area of $\sim$47$\times$47 pc$^2$ around Sgr A$^*$ using the OTF (On the Fly) observing mode, but in this paper we have focused on a region of a $\sim$8$\times$8 pc$^2$ centered at the position of the radio source Sgr A$^*$ ($(\alpha,\delta)_{J2000}$=17$^{\rm h}$45$^{\rm m}$40$^{\rm s}$.031, $-$29$^{\circ}$00$'$28$''$.58).
We have also obtained OTF data of the ortho 1669.904 GHz (1670 GHz) 2$_{12}-1_{01}$ H$_2$O transition, but due to the sensitivity only spectra for three selected positions around Sgr A$^*$ (see Table \ref{table01} and Fig.~\ref{fig3} for their positions) have been extracted in order to study the water excitation.
Since the emission from the GC is very extended, the OTF maps were observed in position switching mode, with the reference observed towards the position $\alpha$=17$^{\rm h}$46$^{\rm m}$10$^{\rm s}$.42, $\delta$=$-$29$^{\circ}$07$'$08$''$.04 (J2000).

We have also obtained single position observations of the ortho 547.676 (548 GHz) 1$_{10}$-1$_{01}$ and para 1101.698 (1102 GHz)
1$_{11}$-0$_{00}$ H$_2^{18}$O transitions towards the three selected positions in the vicinity of Sgr A$^*$, where the 2$_{12}$-1$_{01}$ H$_2$O spectra was extracted from the data cube. 
Table \ref{table001} lists the bands of the HIFI instrument where the H$_2$O and H$_2^{18}$O transitions were observed.
The two CND positions, CND$_1$ and CND$_2$, were observed in the southwest and northeast lobes, respectively, of the CND (see Fig.~\ref{fig6}), and the third single position was observed towards the 20 km s$^{-1}$ cloud. The reference position of the single position observations was the same as in the OTF observations.
The observation dates and IDs of the OTF and single position observations are given in Table \ref{table001}.

The raw data were processed using the version 8 of the HIPE\footnote{HIPE is
a joint development by the {\it Herschel} Science Ground Segment Consortium, consisting of ESA, the NASA {\it Herschel} Science Center, and the HIFI,
PACS and SPIRE consortia.} pipeline to a level 2 product. The baseline subtraction and gridding were done using the GILDAS software package\footnote{http://www.iram.fr/IRAMFR/GILDAS}. 
The data were calibrated using hot/cold black body measurements. The intensity scale is the main beam brightness temperature (T$_{\rm mb}$), obtained using the standard main beam efficiencies ($\eta_{\rm mb}$), $\eta_{\rm mb}$=0.75 for the 1$_{10}-1_{01}$ H$_2$O and H$_2^{18}$O transitions, $\eta_{\rm mb}$=0.74
for the 2$_{02}-1_{11}$ H$_2$O, 1$_{11}-0_{00}$ H$_2$O and H$_2^{18}$O transitions, and $\eta_{\rm mb}$=0.71 for the 2$_{12}-1_{01}$ H$_2$O
transition. We have resampled all spectra to a velocity resolution of 5 km s$^{-1}$ appropriate for the linewidths around 20-100 km s$^{-1}$ observed in GC sources. The half--power beam width (HPBW) at the observed H$_2$O and H$_2^{18}$O frequencies are listed in {\mbox Table \ref{table001}}. 
In our study we have also used SPIRE \citep{Griffin} spectra observed with the SPIRE Short Wavelength (SSW) Spectrometer in February 2011. 
The SPIRE data (the observation ID is 1342214842) were also processed utilizing the HIPE (version 8) pipeline to Level 2.


\section{Maps and spectra of H$_2$O towards the 64 pc$^2$ region around Sgr A$^*$}\label{Results1}

\begin{figure*}
\centering
\includegraphics[width=14.9cm]{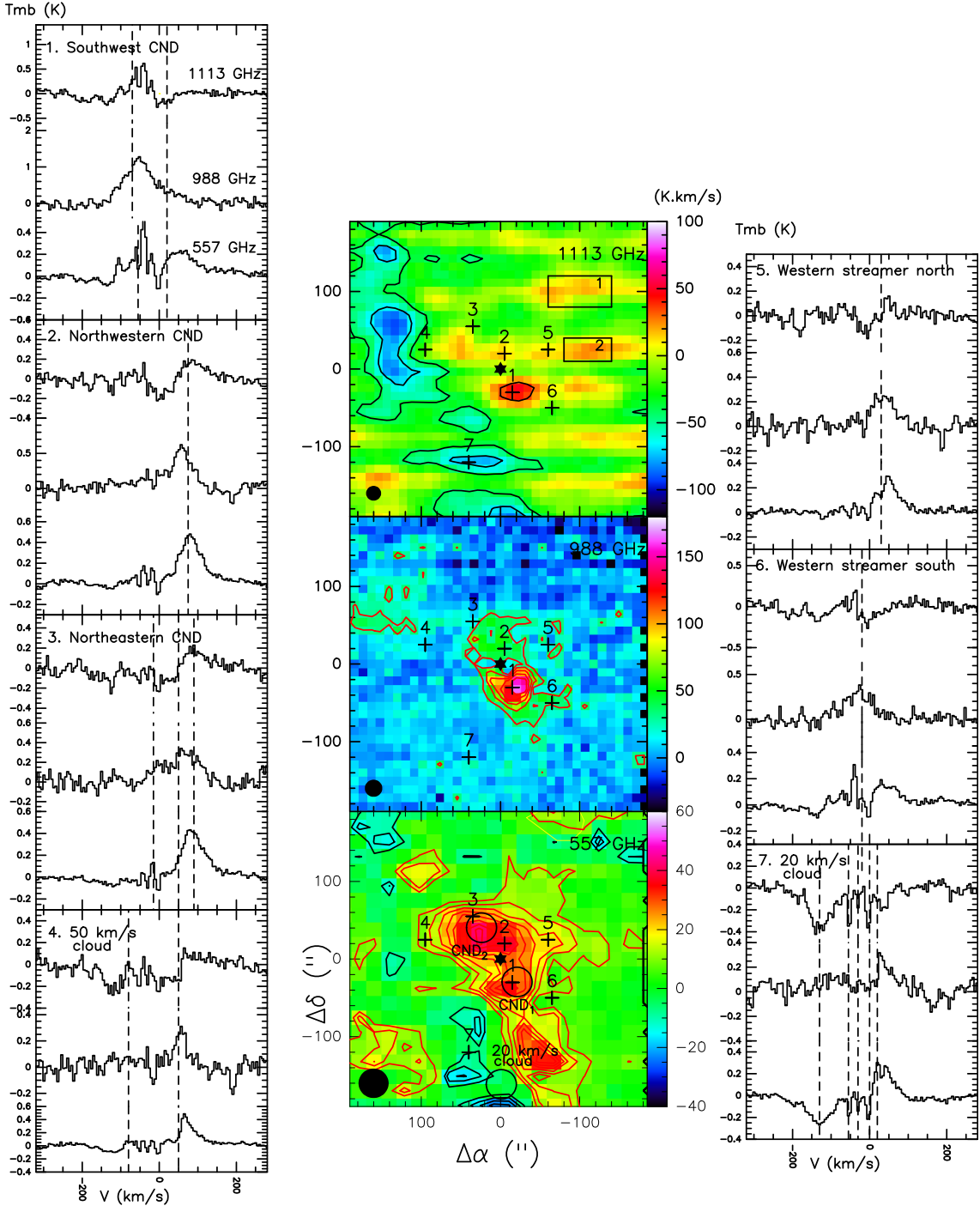}
\caption{{\bf (Central panels)} Velocity integrated intensity maps of H$_2$O at 557, 988 and 1113 GHz. The beam sizes are shown in the left corner of each map. The maps at the three frequencies were 
integrated over the velocity range of [-180,+140] km s$^{-1}$. The first contour levels for the H$_2$O maps at 557 and 1113 GHz are at -3$\sigma$ (black contour) for the absorption, 3$\sigma$ 
(red contour) for the emission, but for the 988 GHz line is at 3$\sigma$ (red contour) for the emission. The steps are of 4.5 K 
km s$^{-1}$ (-4.5 K km s$^{-1}$ for the absorption) at 557 GHz and 26.0 K km s$^{-1}$ (-26.0 K km s$^{-1}$ for the absorption
at 1113 GHz) at 988 and 1113 GHz ($\sigma$=4.1, 11.3 and 9.6 K km s$^{-1}$ at 557, 987 and 1113 
GHz, respectively). Sgr A$^*$ is shown with a black star and it is the origin of the offsets. Black crosses and their numbers show the positions where spectra of the left and right panels were extracted. Every position is associated with a representative GC feature (see below). The three positions where the H$_2^{18}$O spectra were obtained are also shown with black circles on the 557 GHz water map and labeled as CND$_1$, CND$_2$ and 20 km s$^{-1}$ cloud (see Fig.~\ref{fig8}); The H$_2$O and H$_2^{18}$O emission toward these positions are modeled in Sec. \ref{Model}. The wedges to the right show the H$_2$O integrated intensity scale. The parallelograms 1 and 2 shown in the 1113 GHz H$_2$O map were used to extract the average spectra indicated in Fig.~\ref{standing_wave}, illustrating the baseline levels. 
{\bf (Left and Right panels)} Spectra from 7 positions of the 557, 988 and of the 1113 GHz lines. 
Each spectrum was extracted from H$_2$O cubes convolved with the HIFI 38$\arcsec$ beam of the 557 GHz line. The numbers in the upper left side of each subpanel show positions (indicated in the H$_2$O maps) associated with GC features. These features and their systemic velocities (indicated with dashed lines in the H$_2$O spectra) are: 1) the southwest CND (-70 
km s$^{-1}$, this position also covers the northern part of the 20 km s$^{-1}$ cloud),
2) the northwestern CND (75 km s$^{-1}$), 3) the northeastern CND (90 km s$^{-1}$, this position also covers the Northern Ridge with a velocity of -15 km s$^{-1}$ and the 50 km s$^{-1}$ cloud), 4) the 50 km s$^{-1}$ cloud (this position also covers Cloud A with a velocity of -80 km s$^{-1}$), 5) the Western streamer north (30 km s$^{-1}$), 6) the Western streamer south (-20 km s$^{-1}$) and 7) the 20 km s$^{-1}$ cloud. In the subpanel of position 7 we have also indicated with dashed lines the velocities of the absorption features associated with foreground sources (see Sec. \ref{Results1}).}
\label{fig3}
\end{figure*}

The central panels in Fig.~\ref{fig3} show the integrated intensity maps of the three H$_2$O transitions at 557, 988 and 1113 GHz.
The three maps were obtained by integrating over the velocity range between -180 km s$^{-1}$ and +140 km 
s$^{-1}$.  
In Fig.~\ref{fig3} we observe H$_2$O emission/absorption features in the 557 and 1113 GHz maps, whereas only emission dominates the 988 GHz map. Unfortunately, the H$_2$O map at 1113 GHz is affected by striping along the scanning direction due to standing waves, which originate from the Local Oscillator feed horns of HIFI as described in the HIFI handbook\footnote{http://herschel.esac.esa.int/twiki/pub/HSC/HiFi/hifi\_handbook.pdf}.
Our 1113 GHz data were affected by standing waves as the 1113 GHz water transition 
falls at the edge of the mixer band 4, where the standing waves are more prominent (see Sec. 5.3 in the HIFI handbook).  
We have not been able to remove these standing waves in our data with our baseline subtraction and even applying the methods recommended in the HIFI data reduction guide\footnote{http://herschel.esac.esa.in/hcss-doc-15.0/load/hifi\_um/html/hifi\_um.html}. Average spectra at 1113 GHz with standing waves (with amplitudes around 0.15 K) are shown in Fig.~\ref{standing_wave} (Appendix \ref{Standingwave}), which have been extracted over parallelograms 1 and 2 drawn in the 1113 GHz H$_2$O map shown in Fig.~\ref{fig3}.

To study the physical conditions and the chemical composition of GC features, \citet{Baladron} used 7 representative positions selected from
their SiO(2--1) emission maps. The positions are associated with the following features: 1) the southwest CND (this position also covers the northern part of the 20 km s$^{-1}$ cloud), 2) the northwestern CND, 3) the northeastern CND (this position also covers the 50 km s$^{-1}$ cloud and the northern Ridge), 4) the 50 km s$^{-1}$ cloud (this position also covers Cloud A), 5) the western streamer north, 6) the western streamer south, and 7) the 20 km s$^{-1}$ cloud. Fig.~\ref{fig3} also shows line profiles of ortho 1$_{10}-1_{01}$, and para 2$_{02}-1_{11}$ and 1$_{11}-0_{00}$ H$_2$O transitions extracted from the previous 7 positions. 
All spectra at the three frequencies were extracted from H$_2$O cubes convolved to the 38$\arcsec$ beam of HIFI at 557 GHz.

\begin{sidewaysfigure}
\includegraphics[scale=0.48,angle=0]{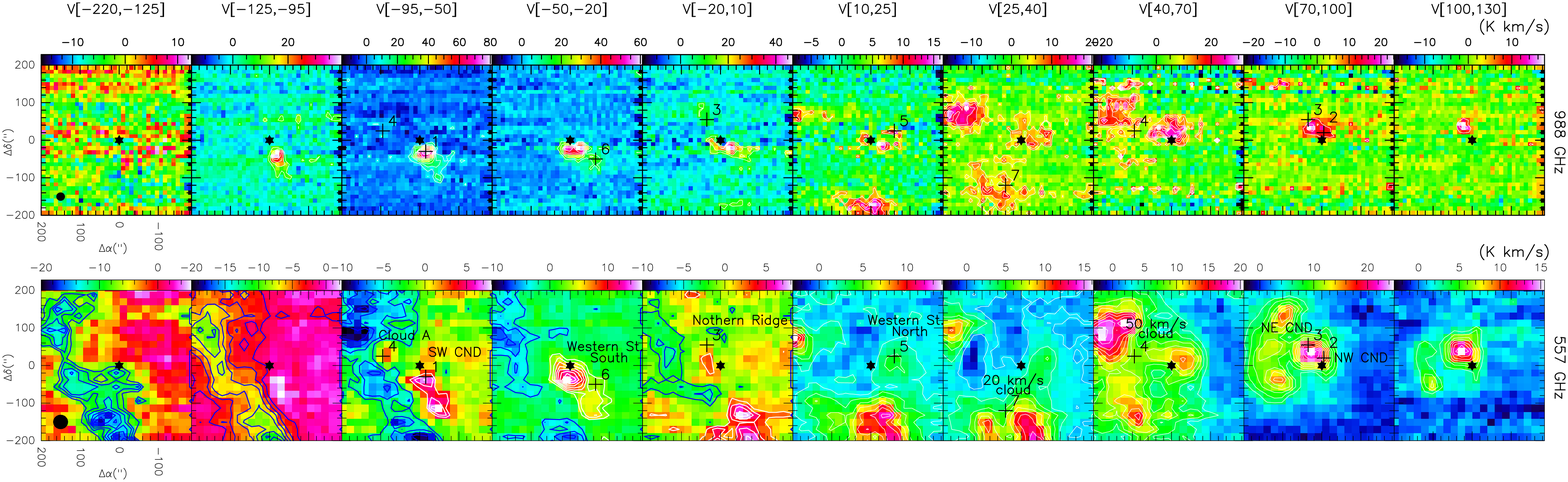}
\caption{Integrated intensity maps of H$_2$O at 988 GHz (first row) and 557 GHz (second row).
The velocity ranges are indicated at the top of each column. The first contour levels of H$_2$O (557 GHz) are at -3$\sigma$ (blue contour) for
the absorption, 3$\sigma$ (white contour) for the emission in steps of 1.5 K km s$^{-1}$ (-1.5 K km s$^{-1}$ for the absorption, 
$\sigma$ in the range 1.0--1.5 K km s$^{-1}$ for all velocity ranges). The first contour levels of H$_2$O (988 GHz) are at 3$\sigma$ (white contour) for the emission 
in steps of 7 K km s$^{-1}$ ($\sigma$ in the range 2.0--3.5 K km s$^{-1}$ for all velocity ranges).
The wedge above each panel shows the H$_2$O integrated intensity scale given in K km s$^{-1}$.
The black star represents Sgr A$^*$ and the origin of the offsets in arcsec. Black crosses and their numbers show positions associated with GC sources labeled in the H$_2$O maps at 557 GHz. Spectra from those positions are shown in Fig.~\ref{fig3}. Beam sizes (38$\arcsec$ at 557 GHz and 22$\arcsec$ at 988 GHz) are shown in the left corner of the first column.}
\label{fig4}
\end{sidewaysfigure}

To identify the emission/absorption from GC features in Fig.~\ref{fig4} we have shown the spatial distribution of the water emission/absorption in the ortho 1$_{10}-1_{01}$ and para 2$_{02}-1_{11}$ H$_2$O transitions integrated over 10 velocity ranges. In this figure, para 1$_{11}$-0$_{00}$ H$_2$O maps are not shown as the data cube suffers from standing waves, causing striping in the 1113 H$_2$O maps. The 557 and 988 GHz maps are only slightly affected by striping (see Fig.~\ref{fig4}). We have used the same velocity ranges in our velocity integrated intensity H$_2$O maps as those used by \citet{Baladron}.
In Fig.~\ref{fig4} the black crosses show the same positions associated with GC features as in Fig.~\ref{fig3}, where H$_2$O spectra were extracted from.

\begin{sidewaysfigure}
\includegraphics[scale=0.53,angle=0]{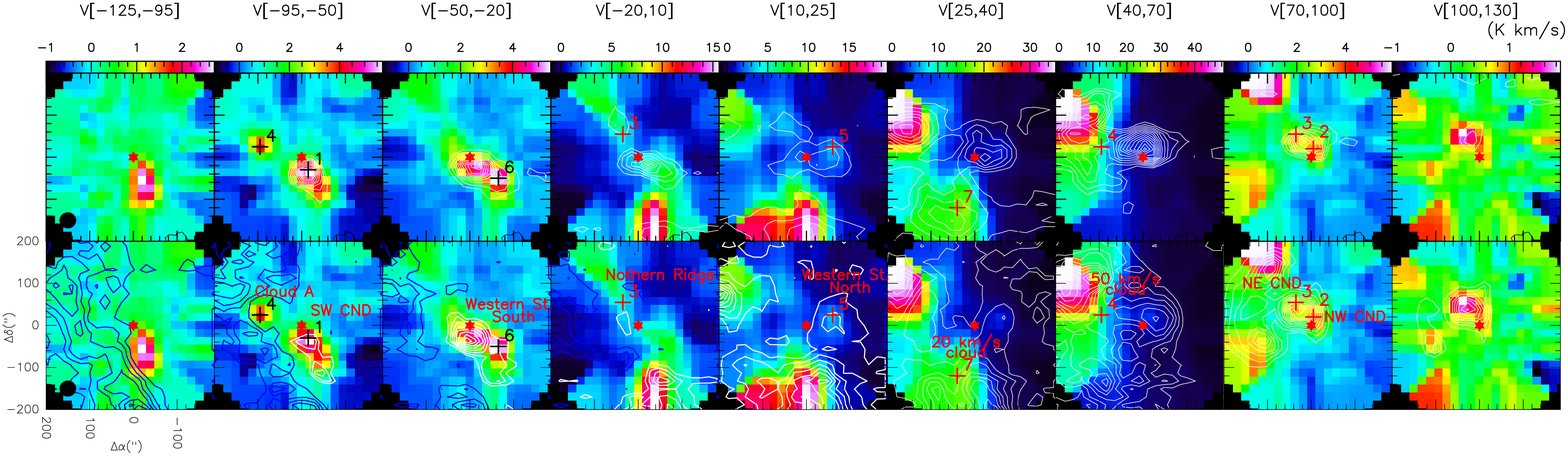}
\caption{Comparison between the SiO(2--1) maps (background images) obtained by \citet{Baladron} and our H$_2$O maps at 988 GHz (first row) and 557 GHz (second row). The velocity ranges are indicated at the top of each column. The wedges at the top of the first row show the SiO(2--1) intensity gray scale in K km s$^{-1}$. H$_2$O contour levels (in blue and white for the absorption and emission, respectively) start at 3$\sigma$ (-3$\sigma$) for the emission 
(absorption) at the two H$_2$O frequencies, and they increase in 5$\sigma$ (-5$\sigma$) and 4$\sigma$ steps 
at 557 and 988 GHz, respectively. The red star represents Sgr A$^*$ and origin of the offsets in arcsec. Crosses and their numbers show positions associated with GC sources labeled in the second row. H$_2$O spectra from those positions are shown in Fig.~\ref{fig3}. The water and SiO(2--1) maps have the same beam size of 38$\arcsec$ shown in the left corner of the first column.}
\label{fig5}
\end{sidewaysfigure}

In Fig.~\ref{fig5} we have compared our ortho 1$_{10}-1_{01}$ and para 2$_{02}-1_{11}$ H$_2$O maps with J=2-1 emission maps of the shock tracer
SiO \citep{Pinta97}, obtained by \citet{Baladron}. For the comparison, previously the 988 GHz H$_2$O and SiO(2--1) maps were convolved to the 38 
$\arcsec$ beam of the ortho 557 GHz H$_2$O map.

\subsection{Analysis of the H$_2$O spectra}\label{Results3}

We have noted in Fig.~\ref{fig3} that the ortho 1$_{10}-1_{01}$ and para 1$_{11}$-0$_{00}$ H$_2$O lines are absorption--dominated in almost all positions, while the para 2$_{02}-1_{11}$ H$_2$O lines are emission--dominated in all positions.
Most spectra from the ground state ortho 1$_{10}-1_{01}$ and 
para 1$_{11}-0_{00}$ H$_2$O transitions reveal the presence of narrow absorption features at V$_{LSR}$=0, $-$30,$-$55
km s$^{-1}$, as well as a broad absorption feature at $\sim$ $-$130 km s$^{-1}$. The absorption features at 0, $-$30 and $-$55 km s$^{-1}$ have been associated with the Local Sgr, $-$30 km s$^{-1}$ and 3--kpc Galactic Arms, respectively, and the broad absorption at -130 km s$^{-1}$ with the Molecular Ring located $\sim$180 pc around the GC \citep{Sandqvist}.

The CND and the 20 km s$^{-1}$ cloud have been studied by \citet{Baladron} using the emission from SiO, H$^{13}$CO$^+$, HN$^{13}$C, HNCO, C$^{18}$O and CS.
Based on that study, we expect that the water emission towards the CND$_1$ and CND$_2$ positions could be affected by water emission/absorption
from the 20 and 50 km s$^{-1}$ clouds in the velocity ranges of $\sim$[-10,+40] and $\sim$[+10,+70] km s$^{-1}$, respectively, whereas the 20 km s$^{-1}$ cloud position is not expected to be affected by water emission/absorption from any positive--velocity source along this line of sight. These three positions are indicated in Fig.~\ref{fig3}. In fact, we have seen in Fig~\ref{fig3} that the 
ortho 1$_{10}-1_{01}$ and para 2$_{02}-1_{11}$ water line profiles of positions 2 and 3, which are close to our CND$_2$ position, reveal signs of contribution in the water emission from the 50 km s$^{-1}$ cloud. We have also noted in Fig.~\ref{fig3} that the water line profiles of position 1, that coincides with our CND$_1$ position, would be affected by water emission/absorption from the 20 km s$^{-1}$ cloud. Additionally, we have seen in Fig.~\ref{fig3} that the water line profiles of position 7, located around 57$\arcsec$ northeast from our 20 km s$^{-1}$ cloud position, are not affected by water emission/absorption from other positive--velocity line of sight sources.

\subsection{Analysis of the 1$_{10}-1_{01}$ H$_2$O emission/absorption distribution towards GC sources}\label{Map557}

As seen in Fig.~\ref{fig4}, at 557 GHz absorption features are observed from -220 km s$^{-1}$ to 10 km s$^{-1}$. The emission at 557 GHz covers the velocity range [-95,130] km s$^{-1}$. The absorption features at 557 GHz within the velocity range of [-220,+10] km s$^{-1}$ correspond to the Local Sgr, -30 km s$^{-1}$, 3--kpc Arms and the Molecular Ring. 
We have found that the ortho 1$_{10}-1_{01}$ H$_2$O emission peaks in the extreme blue--shifted velocity range of [-95,-20] km s$^{-1}$ in the southwest CND, as well as at the extreme red--shifted velocity
range of [+70,+130] km s$^{-1}$ in the northwestern and the northeastern CND.
Ortho 1$_{10}-1_{01}$ H$_2$O emission is not detected in Cloud A at the velocity range of [-95,-50] km s$^{-1}$ due to likely the 3-kpc arm absorption (see spectra of position 4 in Fig.~\ref{fig3}). In the velocity ranges of [+25,+40] and [+40,+70] km s$^{-1}$ we have detected ortho 1$_{10}-1_{01}$ H$_2$O emission from the 20 and 50 km s$^{-1}$ clouds, respectively.
We have also detected ortho 1$_{10}-1_{01}$ H$_2$O emission towards the Western streamer south and Western streamer north in the 
velocity ranges of [-50,-20] and [+10,+25] km s$^{-1}$, respectively. Ortho 1$_{10}-1_{01}$ H$_2$O emission is not detected in the Northern Ridge at the velocity range of [-20,+10] km s$^{-1}$, probably due to the absorption at $\sim$0 km s$^{-1}$ by
the Local Sgr Arm (see spectra of position 3 in Fig.~\ref{fig3}).  

Moreover, Fig.~\ref{fig5} shows a very good agreement between the emission of SiO(2--1) and ortho 1$_{10}-1_{01}$ H$_2$O in 
the CND, Western Streamer, and the 20 km s$^{-1}$ and 50 km s$^{-1}$ clouds.

\subsection{Analysis of the 2$_{02}-1_{11}$ H$_2$O emission distribution towards GC sources}\label{Map988}

\begin{figure}
\includegraphics[scale=0.37]{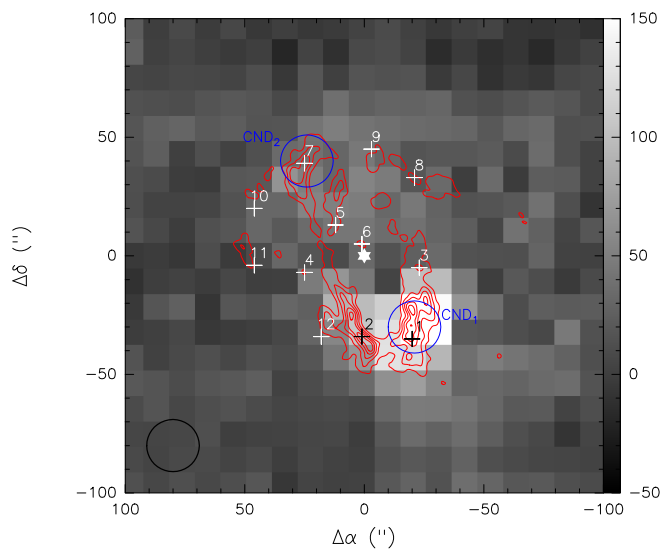}
\caption{Comparison between the integrated intensity maps of 988 GHz H$_2$O in gray and CN(2--1) (contours,
\citet{Martin}). The integrated velocity range of the H$_2$O map is [-180,140] km s$^{-1}$. The wedge at the right shows the H$_2$O intensity gray scale in K km s$^{-1}$. Crosses and their numbers show positions, where spectra used in the kinematic study were extracted over the 22$\arcsec$ beam. These positions correspond to selected CN(2--1) and H$_2$CO(3$_{30}$--2$_{20}$) emission peaks on the CND \citep{Martin}. The white star shows the position of Sgr A$^*$ and origin of the offsets. The HIFI beam of 22$\arcsec$ at 988 GHz is shown in the left corner. Contour levels of the CN map start at 3$\sigma$ and increase in 4$\sigma$ steps. 22$\arcsec$ blue circles show the CND$_1$ and CND$_2$ positions selected for our study of the water excitation (see Section \ref{Model}). The CND$_1$ and CND$_2$ positions were observed towards the southwest and northeast lobes, respectively, of the CND \citep{Reque12}.}
\label{fig6}%
\end{figure}

As mentioned above the para 988 GHz H$_2$O line only shows emission (see Fig.~\ref{fig3} and~\ref{fig4}). This 
emission is concentrated in all previously mentioned GC features except in Cloud A and the Northern Ridge. Roughly the para 2$_{02}-1_{11}$ H$_2$O emission exhibits a good 
correlation with the SiO(2--1) emission arising from GC sources in the vicinity of Sgr A$^*$ (see Fig.~\ref{fig5}). 
In Fig.~\ref{fig6} we have compared the interferometric map of CN(2--1) \citep{Martin} 
with the para 2$_{02}-1_{11}$ H$_2$O emission map of the CND. Despite the difference in the spatial resolution between both maps, it can be clearly seen that there is an excellent correlation between the emission of the CN(2--1) and the para 
2$_{02}-1_{11}$ H$_2$O emission towards the southwest CND.

\subsection{H$_2$O and H$_2^{18}$O detection in selected GC positions}\label{H218O_detection}
\begin{figure*}
\centering
\includegraphics[scale=0.8]{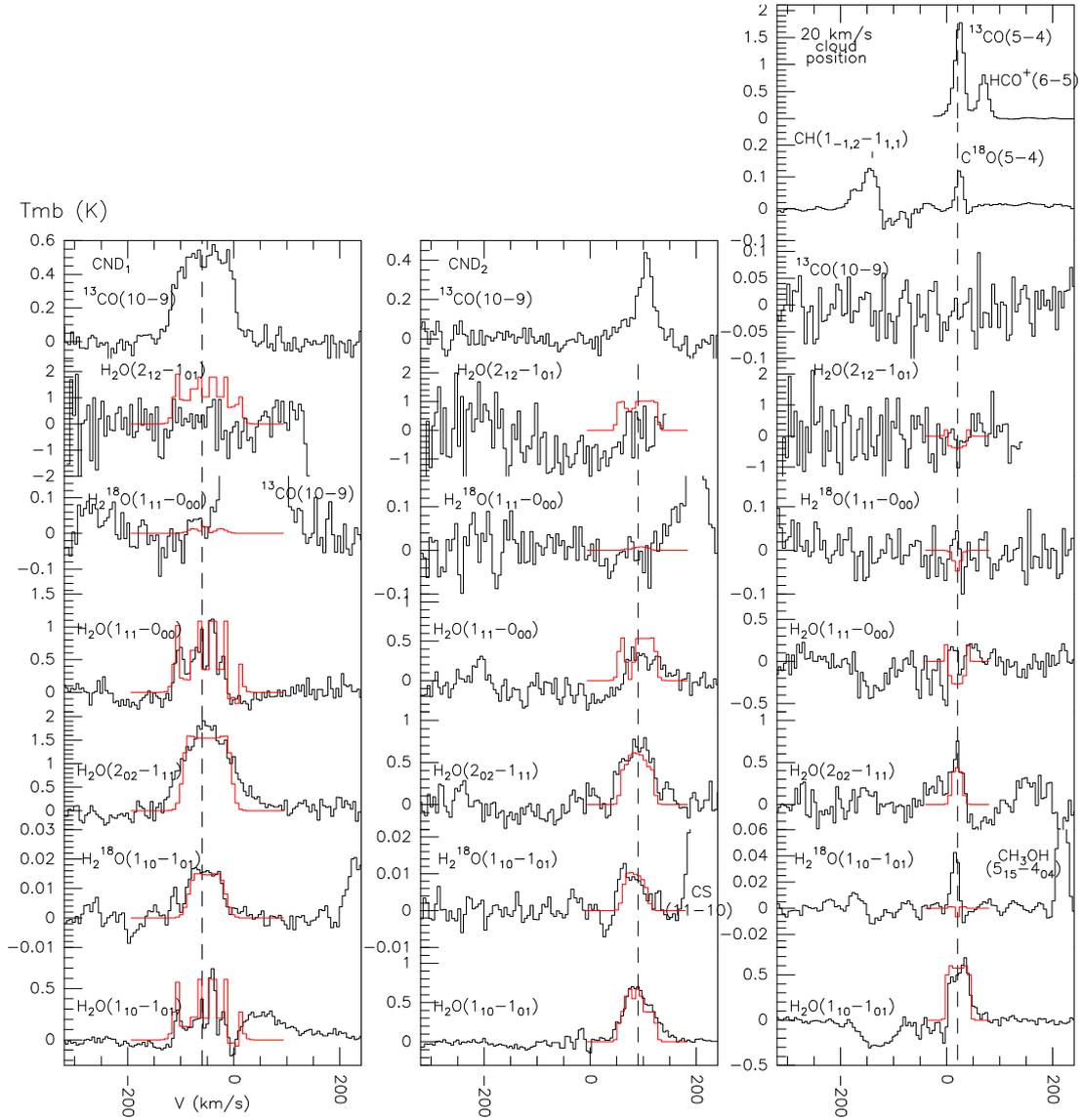}
\caption{Observational and simulated H$_2$O and H$_2^{18}$O line profiles. 
{\bf (Left panel)} Observational spectra (black histograms) of 4 and 2 transitions of H$_2$O and H$_2^{18}$O, respectively, for the CND$_1$ position. The modeled water line profiles obtained using a two component model are shown with red histograms. We also show the $^{13}$CO(10-9) line profile. {\bf (Central panel)} Observational spectra (black histograms) of 4 and 2 transitions of H$_2$O and H$_2^{18}$O, respectively, for the CND$_2$ position. The modeled water line profiles obtained utilizing a two component model are shown with red histograms. $^{13}$CO(10-9) and CS(11-10) line profiles are also shown, but the CS(11-10) line (near the H$_2^{18}$O(1$_{10}-1_{01}$) line) appears as a consequence of the HIFI double sideband. {\bf (Right panel)} Observational spectra (black histograms) of 4 and 2 transitions of H$_2$O and H$_2^{18}$O, respectively, for the 20 km s$^{-1}$ cloud position. The modeled water line profiles are shown with red histograms. $^{13}$CO(10--9), 
C$^{18}$O(5--4), $^{13}$CO(5--4), CH$_3$OH(5$_{15}-4_{04}$), CH(1$_{-1,2}-1_{1,1}$) and HCO$^+$(6--5) spectral features are also shown, with the last three lines coming from the HIFI double sideband. The H$_2^{18}$O(1$_{10}-1_{01}$) line is blended with the $^{13}$CH$_3$OH(1$_{62}-1_{61}$) line (this line also is due to the HIFI double sideband).}
\label{fig8}
\end{figure*}

In Fig.~\ref{fig8} we show the H$_2^{18}$O spectra, as well as the H$_2$O spectra extracted from the data cubes for the CND$_1$, CND$_2$ and 20 km s$^{-1}$ cloud
positions indicated in Fig.~\ref{fig3}. 
We have detected emission of the ortho 1$_{10}$-1$_{01}$ H$_2^{18}$O transition towards the CND, but unfortunately this transition is blended with the 
$^{13}$CH$_3$OH(1$_{62}-1_{61}$) line (this line is due to the HIFI double sideband) for the 20 km s$^{-1}$ cloud position. Moreover, emission/absorption of the 
para 1$_{11}-0_{00}$ H$_2^{18}$O transition for the three studied positions was not detected with our sensitivity. Other molecular spectral features were found in the spectra of 
ortho 1$_{10}$-1$_{01}$ and para 1$_{11}-0_{00}$ H$_2^{18}$O transitions (see Fig.~\ref{fig8}). 
Given our spectral sensitivity, the emission/absorption of the ortho 2$_{12}-1_{01}$ H$_2$O 
transition (unfortunately this transition falls in the edge of the observed band) was not detected towards the CND and 20 km s$^{-1}$ cloud positions, while the ortho 
1$_{10}-1_{01}$, and para 2$_{02}-1_{11}$ and 1$_{11}-0_{00}$ H$_2$O transitions reveal emission/absorption for the three studied positions.
These H$_2$O and H$_2^{18}$O spectra will be used in our study of the water excitation in Sec. \ref{Model}.

\begin{figure}
\centering
\includegraphics[width=9cm]{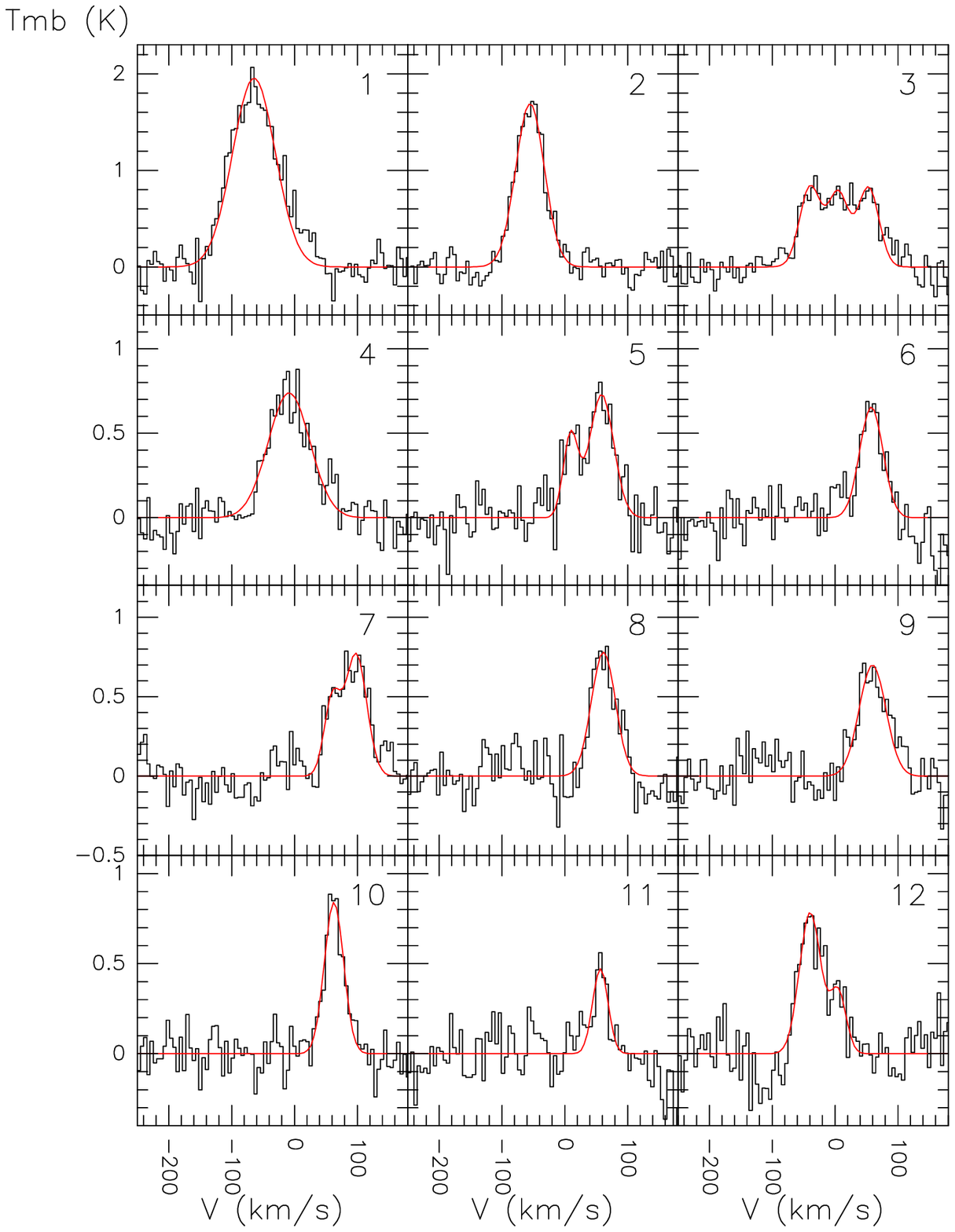}
\caption{Water 2$_{02}$-1$_{11}$ line profiles (black histogram) observed towards 12 positions (indicated in Fig.~\ref{fig6}) on the CND. Gaussian fits to the water lines are shown with a red line. To fit the water line profiles we have used a single Gaussian except in positions 3, 5, 7 and 12 (three Gaussians in position 3 and two Gaussians in positions 5, 7 and 12).}
\label{Appen_fig1}
\end{figure}

\section{Water Kinematics}\label{Kinematic}
\begin{figure}
\includegraphics[scale=0.7]{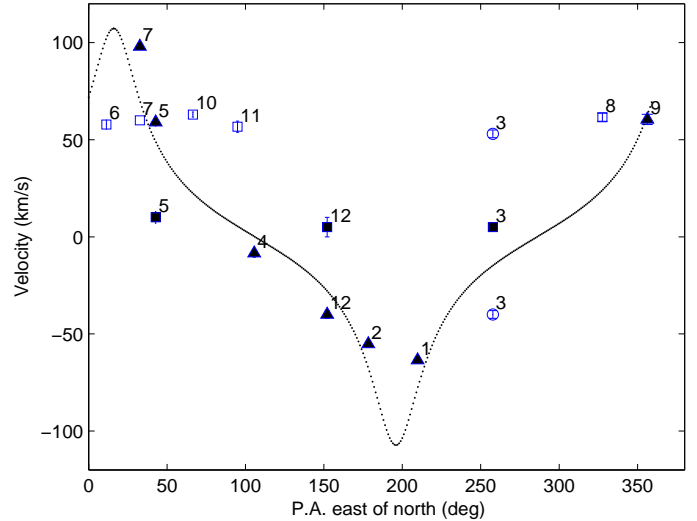}
\caption{LSR velocities of 988 GHz H$_2$O lines for 12 positions on the CND (see Fig.~\ref{fig6}) represented as function of the PA. Different symbols represent several sources in the H$_2$O map at 988 GHz: the CND (filled triangles),
the 50 km s$^{-1}$ cloud (open squares), the 20 km s$^{-1}$ cloud (filled squares) and the Western Streamer (open circles). There are two velocity components in positions 5, 7 and 12, and three velocity components in position 3 (see text). The black dotted line represents our best fit of a rotating ring model to positions 1, 2, 4, 5, 7, 9 and 12. The error bars correspond to LSR 
velocity errors in the Gaussian fits. Error bars overlap with some symbols. The 5 km s$^{-1}$ velocity component in position 3 and both velocity components in position 7 do not have error bars as the velocity was fixed in the Gaussian fit.}
\label{fig7}%
\end{figure}

We have studied the CND kinematics of water vapor using the para H$_2$O emission at 988 GHz since it is the least affected by absorption (see Fig.~\ref{fig3} and~\ref{fig4}). This transition also exhibits significant emission arising from the Western Streamer, and the 20 and 50 km s$^{-1}$ clouds as seen in Fig.~\ref{fig4}. To derive the kinematics we have selected 2$_{20}-1_{11}$ H$_2$O spectra (see Fig.~\ref{Appen_fig1}) corresponding to the CN(2--1) and H$_2$CO(3$_{03}-2_{20}$) emission peaks studied by \citet{Martin}.
They found that the CN(2--1) emission is an excellent tracer of the CND, while the H$_2$CO(3$_{03}-2_{20}$) emission traces a shell--like structure where Sgr A East and both clouds seem to be interacting. Fig.~\ref{fig6} shows the selected positions for the kinematic study superimposed on the para 2$_{20}-1_{11}$ H$_2$O map. Positions 1--8 correspond to CN(2--1) emission peaks, while positions 9--12 correspond to H$_2$CO(3$_{03}-2_{20}$) emission peaks. Gaussian fits 
to the water para 2$_{20}-1_{11}$ line profiles were performed (see Fig.~\ref{Appen_fig1}) and the derived parameters are shown in Table~\ref{table1}.
The LSR (Local Standard of Rest) velocities as a function of Position Angles (PA) are represented in Fig.~\ref{fig7}. 
The PA is measured east from north centered on Sgr A$^*$. 

To describe the water kinematics from the CND, Fig.~\ref{fig7} shows the model prediction of the LSR velocities of a rotating ring model with an inclination of 75$^\circ$, a position angle of 196$^\circ$ and a rotation velocity $v_{rot}$/$sin\,i$=115 km s$^{-1}$ of our best fit
for the CND components (triangles).
Our derived inclination angle, PA and rotation velocity are slightly lower than those estimated by \cite{Martin} for the southwest lobe of the CND. Our derived three parameters are in agreement with those of rotating rings used to model the CND kinematics in \cite{Goicoe2018a}.   

Limited by the {\it Herschel} spatial resolution, Fig. \ref{fig7} shows the presence of 4 kinematically distinct structures around Sgr A$^*$: the CND represented by filled triangles, the 50 
km s$^{-1}$ cloud represented by open squares, the 20 km s$^{-1}$ cloud indicated by filled squares, and the Western Streamer represented by open circles. The 50 and 20 km s$^{-1}$ clouds and the Western Streamer are located on top of the CND \citep{Martin}. The presence of more than one velocity component towards position 3, 5, 7 and 12 can be clearly seen in the 988 GHz spectra shown in Fig.~\ref{Appen_fig1}. The velocity components of -40 and 53 km s$^{-1}$ observed in position 3 are consistent with those arising from the southern and northern parts, respectively, of the Western Streamer \citep{Baladron}.
As shown in Fig.~\ref{fig7}, there are three features not described by the rotating ring model, the 20 km s$^{-1}$ and 50 km s$^{-1}$ clouds and the Western Streamer.
However, the 60 km s$^{-1}$ velocity component of position 7 could be associated with the CND rather than with the 50 km s$^{-1}$ cloud.
Our result is consistent with that of \citet{Martin}, who found kinematically distinct 
features in the vicinity of the CND using the CN(2--1) emission and indicating that the water emission traces both components, the CND and the clouds 
interacting with the SNR Sgr A East.

\begin{table}
\caption{Gaussian fit parameters of 998 GHz H$_2$O lines on selected positions of the CND.}
\begin{tabular}{ccrrrc}
\hline
\hline
Pos. & $\Delta\alpha$,$\Delta\delta$\tablefootmark{a} & Area$\pm\sigma$ & V$_{LSR}$$\pm\sigma$ & $\bigtriangleup_{v_{1/2}}\pm\sigma$ & T$_{mb}$\\
 & (\arcsec,\arcsec) & (K km s$^{-1}$) & (km s$^{-1}$) & (km s$^{-1}$) & (K)\\
\hline 
1  & (-26,-35) & 167$\pm$3 & -64$\pm$1 & 80\tablefootmark{b}  & 2.0\\
2  &   (1,-34) & 97$\pm$3  & -55$\pm$1 & 54$\pm$2  & 1.7\\
3  & (-23,-5)  & 34$\pm$3 & -40$\pm$2 & 39\tablefootmark{b}& 0.8\\
   &      & 32$\pm$3 & 5\tablefootmark{b} & 39\tablefootmark{b} & 0.8\\
   &      & 34$\pm$3 & 53$\pm$2 & 39\tablefootmark{b} & 0.8\\
4  & (25,-7)   & 62$\pm$3  & -8$\pm$2  & 79$\pm$5  & 0.7\\
5  & (12,13)   & 14$\pm$4 & 10$\pm$3 & 28$\pm$6 & 0.5\\
   &           & 35$\pm$4 & 59$\pm$2 & 45$\pm$6 & 0.7\\       
6  & (1,5)     & 30$\pm$4  & 58$\pm$3  & 44$\pm$7  & 0.7\\
7  & (25,39)&15$\pm$2 & 60\tablefootmark{b} &30\tablefootmark{b} & 0.5\\
 & & 33$\pm$2&98\tablefootmark{b}&40\tablefootmark{b}& 0.8\\
8  & (-21,33)  & 37$\pm$3  & 62$\pm$2  & 45$\pm$4  & 0.8\\
9  & (-3,45)   & 37$\pm$3  & 60$\pm$2  & 50\tablefootmark{b}  & 0.7\\
10 & (46,20)   & 33$\pm$3  & 63$\pm$1  & 36$\pm$4  & 0.8\\
11 & (46,-4)   & 15$\pm$4  & 57$\pm$3  & 29$\pm$12 & 0.5\\
12 & (18,-34)  & 33$\pm$3  & -40$\pm$2 & 40\tablefootmark{b}  & 0.8\\
   &           & 11$\pm$2 & 5$\pm$5 & 30\tablefootmark{b} & 0.3\\
\hline
\end{tabular}
\tablefoot{
\tablefoottext{a}{Offsets relative to the Sgr A$^*$ position.} \tablefoottext{b}{This value is written without errors as this parameter was fixed in the Gaussian fit.}}
\label{table1}
\end{table}

\begin{figure}
\includegraphics[width=8cm]{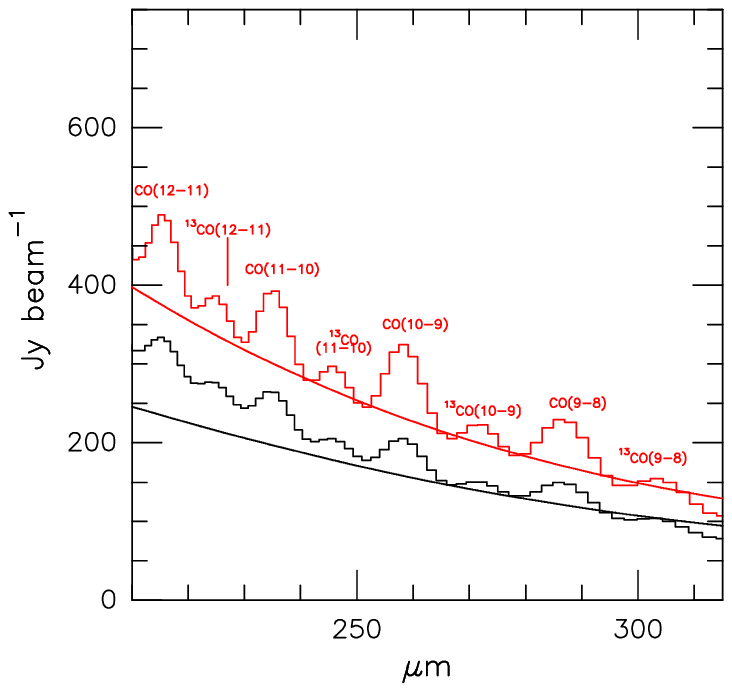}
\caption{Observational SPIRE spectra. The red and black histograms represent the SPIRE spectra for the CND$_1$ and CND$_2$ 
positions, respectively. The molecular lines in the SPIRE spectra correspond to CO and $^{13}$CO lines. The red and black lines show the predicted dust continuum for the CND$_1$ and CND$_2$ 
positions, respectively, obtained using a two component model (see text).}
\label{fig9}
\end{figure}

\section{Modeling the H$_2$O and H$_2^{18}$O spectra and the continuum using a non-local radiative transfer code}\label{Model}
We have used a non--local radiative transfer code \citep{Gonzalez} to model the H$_2$O and H$_2^{18}$O line profiles observed with HIFI for the CND$_1$, CND$_2$ and the 20 km s$^{-1}$ cloud positions, as well as to predict the infrared continuum observed with SPIRE for both CND 
positions.
The numerical code solves the non--LTE equations of statistical equilibrium and radiative transfer in spherical geometry with an assumed radius. The sphere is divided into a set of shells defined by the number of the radial grid points. The physical conditions (water abundance X$_{\rm H_2O}$, H$_2$ density n$_{\rm H_2}$, kinetic temperature T$_{\rm k}$, dust temperature T$_{\rm d}$, microturbulent velocity V$_{\rm t}$ and dust--to--gas mass ratio D/G) are defined in each shell of the spherical cloud, and we have assumed uniform physical conditions for simplicity.
The numerical code convolves
the emerging H$_2$O line intensities to match the angular resolution of the HIFI instrument and the predicted continuum flux is integrated over the source. We have modeled the dust continuum flux by assuming that the source size is equal to the beam size at 250 $\mu$m. In addition to the collisional excitation, the code also accounts for radiative pumping through the dust emission, characterized by the T$_{\rm d}$, the dust opacity $\tau_{\rm d}$ \citep{Gonzalez14} and the D/G.

\begin{table*}
\small
\caption{The lowest $\chi^2$ values for two component model fits to the H$_2$O line intensities}
\centering
\begin{tabular}{ccrcrccccr}
\hline
\hline
 \multicolumn{5}{c}{CND$_1$ position} & \multicolumn{5}{c}{CND$_2$ position} \\
\multicolumn{2}{c}{low--density} & \multicolumn{2}{c}{high--density} & & \multicolumn{2}{c}{low--density} & \multicolumn{2}{c}{high--density} & \\ 
\multicolumn{2}{c}{component} & \multicolumn{2}{c}{component} & & \multicolumn{2}{c}{component} & \multicolumn{2}{c}{component} & \\
X$_{\rm H_2O}^{\rm ortho}$ & T$_{\rm d}$ & X$_{\rm H_2O}^{\rm ortho}$  & T$_{\rm d}$ & $\chi^2$ & X$_{\rm H_2O}^{\rm ortho}$ & T$_{\rm d}$ & X$_{\rm H_2O}^{\rm ortho}$ & T$_{\rm d}$ & $\chi^2$ \\
($\times$10$^{-6}$) & (K) & ($\times$10$^{-6}$) & (K) &  & ($\times$10$^{-6}$) & (K) & ($\times$10$^{-6}$) & (K) &\\
\hline
3 & 15 & 30 & 30 & 13.0 & \bf{1}\tablefootmark{a} & \bf{15}\tablefootmark{a} & \bf{10}\tablefootmark{a} & \bf{25}\tablefootmark{a} & 11.5\\ 
3 & 20 & 30 & 30 & 11.2 & 1 & 20 & 10  & 25 & 12.5\\
3 & 25 & 30 & 30 & 10.7 & 1 & 25 & 10  & 25 & 16.1\\
3 & 30 & 30 & 30 & 10.4 & 3 & 30 & 0.1 & 25 & 13.1\\
5 & 35 & 10 & 30 & 10.1 & 3 & 35 & 0.1 & 25 & 7.2\\
7 & 40 &  5 & 30 &  6.6 & 3 & 40 & 0.1 & 25 & 6.5\\
\bf{7}\tablefootmark{a} & \bf{45}\tablefootmark{a} & \bf{5}\tablefootmark{a} & \bf{30}\tablefootmark{a} & 4.4 & 3 & 45 & 0.3 & 25 & 12.5\\
9 & 50 & 0.7& 30 & 4.4 & 3 & 50 & 0.3 & 25 & 24.9\\
9 & 55 & 0.5& 30 & 7.9 & 3 & 55 & 0.5 & 25 & 43.1\\
\hline
\end{tabular}
\tablefoot{
\tablefoottext{a}{Values that provide the best fits to both the continuum dust emission and the water line intensities.}
}
\label{table61}
\end{table*}

\begin{table*}
\scriptsize
\caption{Parameters of the models for the three selected positions in the vicinity of Sgr A$^*$}
\begin{tabular}{ccrrrrrrrrrrrrr}
\hline
\hline
Position & & D\tablefootmark{b,c} & D/G & source & $\log$ n$_{\rm H_2}$\tablefootmark{c,d} & T$_{\rm k}$\tablefootmark{c,d} & O/P\tablefootmark{c} & $^{16}$O/$^{18}$O & X$_{\rm H_2O}$\tablefootmark{e} & V$_{\rm t}$\tablefootmark{e} & T$_{\rm d}$\tablefootmark{e} & $\log$ $\tau_{\rm d}$\tablefootmark{f} & $\log$ N$_{\rm H_2}\tablefootmark{g}$ & $\log$ N$_{\rm H_2O}$\tablefootmark{h}\\
 & Com.\tablefootmark{a} & (kpc) & ratio\tablefootmark{c} (\%)& radius\tablefootmark{d} (pc) & (cm$^{-3}$) & (K) & & ratio\tablefootmark{c} & ($\times$10$^{-6}$) &(km s$^{-1}$)&(K)&  &(cm$^{-2}$)&(cm$^{-2}$)\\
\hline
CND$_1$ & 1 & 7.9 & 1 & 0.31 & 4.5 & 200 & 3 & 250 & 9.3$^{+3.7}_{-0.5}$ & 23$\pm$3 & 45$^{+6}_{-3}$ &-2.1$^{+0.04}_{-0.07}$& 22.48$^{+0.02}_{-0.09}$& 17.45$^{+0.15}_{-0.09}$\\
        & 2 & 7.9 & 1 & 0.08 & 5.2 & 500 & 3 & 250 & 6.7$^{+2.0}_{-0.7}$ & 23$\pm$3 & 30$^{+3}_{-3}$ &-2.0$^{+0.04}_{-0.07}$& 22.59$^{+0.01}_{-0.10}$& 17.42$^{+0.11}_{-0.11}$\\
CND$_2$ & 1 & 7.9 & 1 & 0.32 & 4.5 & 175 & 3 & 250 & 1.3$^{+1.2}_{-0.1}$ & 14$\pm$3 & 15$^{+13}_{-13}$ &-2.1$^{+0.04}_{-0.07}$& 22.50$^{+0.01}_{-0.11}$& 16.62$^{+0.28}_{-0.12}$\\
        & 2 & 7.9 & 1 & 0.06 & 5.3 & 325 & 3 & 250 & 13.3$^{+9.3}_{-1.3}$ &14$\pm$3 & 25$^{+3}_{-1}$ &-2.0$^{+0.04}_{-0.07}$& 22.56$^{+0.04}_{-0.08}$& 17.69$^{+0.23}_{-0.09}$\\
20 km s$^{-1}$ cloud & 1 & 7.9 & 1 & 2.3 & 4.6 & 100 & 3 & 250 & 0.04$^{+0.03}_{-0.01}$ & 9$\pm$2 & 26\tablefootmark{i} & -1.1$^{+0.04}_{-0.05}$ & 23.44$^{+0.06}_{-0.06}$ & 16.04$^{+0.25}_{-0.14}$\\
\hline 
\end{tabular}
\tablefoot{
\tablefoottext{a}{Component.}
\tablefoottext{b}{Distance to source.}
\tablefoottext{c}{Fixed parameter.}
\tablefoottext{d}{These values are taken from \cite{Reque12} for the two CND positions. The n$_{\rm H_2}$ and T$_{\rm k}$ are taken from \cite{Baladron} and \cite{Huttemei}, respectively, for the 20 km s$^{-1}$ cloud, while the radius of this source is estimated from our H$_2$O maps (see Appendix \ref{source_size}).}
\tablefoottext{e}{Free parameter.}
\tablefoottext{f}{Predicted value at 250 $\mu$m. The uncertainty in the $\tau_{\rm d}$ is calculated assuming a 10\% uncertainty in the source size.}
\tablefoottext{g}{Estimated N$_{\rm H_2}$ assuming a mass--absorption coefficient of 8.2 cm$^{2}$ g$^{-1}$ at 250 $\mu$m \citep{Gonzalez14} and a D/G of 1\%.}
\tablefoottext{h}{N$_{\rm H_2O}$ derived using the N$_{\rm H_2}$ and X$_{\rm H_2O}$ given in this table.}
\tablefoottext{i}{The T$_{\rm d}$ used for the 20 km s$^{-1}$ position is fixed as no observational continuum was available to constrain the T$_{\rm d}$ of this position.}}
\label{table2}
\end{table*}

We have run two component models to reproduce the H$_2$O and H$_2^{18}$O line profiles for both CND positions using two fixed values of n$_{\rm H_2}$ and T$_{\rm k}$ inferred by \citet{Reque12}, who explained the CO excitation in the CND with two components, one with T$_{\rm k}$ of $\sim$200 K and n$_{\rm H_2}$ of 
$\sim$3.2$\times$10$^4$ cm$^{-3}$ for the low--excitation, and the second with warmer T$_{\rm k}$ of $\sim$300--500 K and n$_{\rm H_2}$ densities of $\sim$2$\times$10$^5$ cm$^{-3}$ for the high--excitation component. The two model components are run separately, and the output line profiles are combined at each position.
The V$_{\rm t}$, T$_{\rm d}$ and X$_{\rm H_2O}$ were considered free parameters and were changed to fit the observed water line profiles, with the T$_{\rm d}$ giving an appropriate fit to both the continuum dust emission and the water line intensities of both CND positions.

For the modeling of the water line profiles for the 20 km s$^{-1}$ cloud position, we have considered a model with fixed n$_{\rm H_2}$ of 4$\times$10$^4$ cm$^{-3}$ \citep{Baladron} and T$_{\rm k}$ of \mbox{100 K} \citep{Huttemei,Rod2001}.
We have adopted a T$_{\rm d}$ of 26 K for the 20 km s$^{-1}$ cloud position as a compromise value between the T$_{\rm d}$ inferred by \citet{Pierce} and \citet{Rod} for the GC. The V$_{\rm t}$ and X$_{\rm H_2O}$ were modified to fit the water lines for the 20 km s$^{-1}$ cloud position.

The V$_{\rm t}$ parameter was fixed in the modeling when a V$_{\rm t}$ value provided the best fit to the H$_2$O line widths. The adopted X$_{\rm H_2O}$ were varied until a good match to the water line intensities observed with {\it Herschel}/HIFI was obtained. When we fix n$_{\rm H_2}$ and vary X$_{\rm H_2O}$, the line intensities depend on the source size. Details on the physical parameters are discussed in Appendix \ref{Conditions}.

It is well know that microturbulent approaches for line formations yield self--absorbed line profiles for optically thick lines (e.g. \cite{Deguchi}), which is not observed in our data even for the very optically thick H$_2$O $1_{10}-1_{01}$ lines. In order to avoid such self--absorption, we have used a coarse grid such that the emergent line shapes of very optically thick lines are flat--top. While this has little effect on the emergent line fluxes (less than 40\% for the H$_2$O 1$_{10}-1_{01}$ line), our modeling is mostly based on the less optically thick H$_2^{18}$O 1$_{10}-1_{01}$ line for which the adopted grid is found to be irrelevant (see below). For all models we have included ten H$_2$O lower rotational--levels.
We have also assumed that the H$_2$O and H$_2^{18}$O molecules have uniform distributions and coexist with dust, and that the ortho to para H$_2$O ratio (O/P) is the typical value of 3. In addition, the H$_2^{18}$O abundance 
relative to H$_2$O is 1/250, according with the $^{16}$O/$^{18}$O isotopic ratio inferred for the GC \citep{Wilson}. In our models the D/G was fixed to the typical value of 1\%.

\subsection{Observational H$_2$O and H$_2^{18}$O spectra and the dust continuum}\label{Data}
As mentioned in Sec.~\ref{H218O_detection} the ortho 1$_{10}-1_{01}$ and para 1$_{11}-0_{00}$ H$_2^{18}$O spectra were taken from the HIFI single position observations in the CND$_1$, CND$_2$ and the 20 km s$^{-1}$ cloud positions, while the ortho 1$_{10}-1_{01}$, 2$_{12}-1_{01}$ and para 2$_{02}-1_{11}$, 1$_{11}-0_{00}$
H$_2$O spectra were extracted for the three positions from the data cubes. These spectra are shown in Fig.~\ref{fig8}. As mentioned in Sec. \ref{Results1}, the H$_2$O spectra at 1113 GHz are affected by standing waves with amplitudes around 0.15 K (see Fig.~\ref{standing_wave}). These artifacts do not affect our analysis of the CND$_1$ position because the intensity of the 1$_{11}-0_{00}$ water line is at least a factor 5 more intense than the amplitude of the artifact. Toward the CND$_2$ and 20 km s$^{-1}$ positions, the observed intensity of the 1$_{11}-0_{00}$ water line could be increased/decreased at most in 0.15 K due to the standing waves, however these possible changes do not affect our best fits for the water line profiles described below.

The SSW SPIRE data were used to extract spectra for the CND$_1$ and CND$_2$ positions. These spectra are shown in Fig.~\ref{fig9}.
Unfortunately, our SPIRE observations did not cover the 20 km s$^{-1}$ cloud position, thus the continuum is not available for this position.
Because of the limited spectral resolution of the SPIRE spectra only CO and $^{13}$CO transitions within J=8--12 can be clearly distinguished in the spectra shown in Fig.~\ref{fig9}.

\subsection{Results}\label{Results2}
We find that, for both the low--density and the high--density model components, the H$_2^{18}$O 1$_{10}-1_{01}$ line is optically thick but effectively optically thin \citep{Snell}, so that its flux can be accurately estimated from $F(\rm erg\, s^{-1} \, cm^{-2})=h \nu n_{\rm H_2}^2 X_{H_2^{18}O}^{ortho} C_{lu} V / 4\pi D^2$, where $C_{lu}$ is the collisional excitation rate from the lower to the upper energy level, $V$ is the volume of the source, and $D$ is the distance. Our best--fit model provides a good match to the line with X$_{H_2^{18}O}^{ortho}$=4$\times$10$^{-8}$ for the high--density component of the CND$_2$ position, which dominates the emission, and also yield fluxes for the H$_2$O lines that are consistent with data. By using the above equation we derive a flux for the H$_2^{18}$O 1$_{10}-1_{01}$ line of 2.6$\times$10$^{-15}$ erg s$^{-1}$ cm$^{-2}$ for the high--density component of the CND$_2$ position, which is only a factor 1.2 higher than that estimated with our model. Such small differences are also found in the cases of the CND$_1$ and the 20 km s$^{-1}$ cloud positions. 
The H$_2$O 1$_{10}-1_{01}$ line is not effectively thin, hence showing a flux significantly weaker than 250$\times$F(ortho H$_2^{18}$O 1$_{10}-1_{01}$).

We have used the $\chi^2$ statistic in order to measure the goodness of fit of the data to the model for both CND positions. In \mbox{Table \ref{table61}} we show the results of a $\chi^2$ test in a two component approach, showing only X$_{\rm H_2O}^{\rm ortho}$ and T$_{\rm d}$ combinations that provide the lowest $\chi^2$ values for model fits to the H$_2$O line intensities. 
The fitting of the dust continuum emission is checked after the $\chi^2$ testing. 
The $\chi^2$ statistic was run for the ranges X$_{\rm H_2O}^{\rm ortho}$=1$\times$10$^{-7}$--9$\times$10$^{-5}$ and T$_{\rm d}$=\mbox{15--55 K}. The derived T$_{\rm d}$ values in bold print in Table \ref{table61} give the best fits to the dust continuum emission of the CND positions (see Figure~\ref{fig9}). These T$_{\rm d}$ values in combination with X$_{\rm H_2O}^{\rm ortho}$ values provide the lowest $\chi^2$ value for the CND$_1$ position, but this does not happen in the case of the CND$_2$ position.
There are two combinations giving a $\chi^2$ value equal to 4.4 in the CND$_1$ position, but that with the inferred T$_{\rm d}$=45 K (for the low--density component) is the one that provides the best fit of the dust continuum. A $\chi^2$ value equal to 6.5 is the lowest value in the CND$_2$ position, but in this case the derived T$_{\rm d}$=40 K of the low--density component overestimates the observed dust continuum. 
Table \ref{table2} summarizes the physical conditions and parameters that provide the best fits of the data.

\subsubsection{CND$_1$ position}
For the CND$_1$ position we have found the best fit for the water line profiles (red histograms in Fig.~\ref{fig8}) with the derived X$_{\rm H_2O}$ of 9.3$\times$10$^{-6}$ and the derived T$_{\rm d}$ of 45 K for the low--density component, and with the derived X$_{\rm H_2O}$ of 6.7$\times$10$^{-6}$ and the derived T$_{\rm d}$ of 30 K for the high--density component. The inferred V$_{\rm t}$ of 23 km s$^{-1}$ provides the best fits to the H$_2$O line widths.

As mentioned above, the observed ortho 1$_{10}-1_{01}$ and para 1$_{11}-0_{00}$ H$_2$O line profiles at negative velocities towards the CND$_1$ position are affected by absorption from foreground sources. To model these absorptions, we have considered spherical shells around the modeled sources, with a water abundance around 3$\times$10$^{-8}$ as derived for the \mbox{$-$30 km s$^{-1}$} spiral arm \citep{Karlsson}, and H$_2$ densities of 10$^{3}$ cm$^{-3}$ and kinetic temperatures of 50 K \citep{Greaves}. Turbulent velocities of \mbox{1--2 km s$^{-1}$} were appropriate to simulate those lines. Water lines are also affected by absorption towards nearby galaxies \citep{Liu}.
The narrow absorption lines superimposed on top of the emission water lines create emission spikes observed in the 1$_{10}-1_{01}$ and 1$_{11}-0_{00}$ H$_2$O lines. There is a very good overall agreement between the observations and the modeling. However, as expected for the complexity of the H$_2$O excitation, there are differences in the intensity of the modeled and the observed spikes likely due to the assumed line--shape in the modeling. This difference is also seen in the case of the H$_2$O 2$_{02}-1_{11}$ line.

Considering an optically thin regime, a dust--to--gas ratio of 1\% and a mass--absorption coefficient of 8.2 cm$^2$ g$^{-1}$ at 250 $\mu$m, the $\tau_{\rm d}$(250 $\mu$m) is related to the H$_2$ column density as N$_{\rm H_2}$/$\tau_{\rm d}$(250 $\mu$m)=3.6$\times$10$^{24}$ cm$^{-2}$. Based on the $\tau_{\rm d}$(250 $\mu$m) predicted with our two component model (see Table \ref{table2}), we have derived a H$_2$ column density N$_{\rm H_2}$ of 3.0--3.9$\times$10$^{22}$ cm$^{-2}$ for the CND$_1$ position, values listed in Table \ref{table2}, together with the water abundances, turbulent velocities, dust temperatures and dust opacities. For this model we have also derived H$_2$O column densities (N$_{\rm H_2O}$) around 3$\times$10$^{17}$ cm$^{-2}$ also included in Table \ref{table2}. A total gas mass of 238 M$_{\odot}$ is derived for the CND$_1$ position by considering the N$_{\rm H_2O}$ and the source size. This mass is a factor 1.6 lower than that estimated in \cite{Reque12} for this position, which is reasonable given the simplicity of our modeling.

\subsubsection{CND$_2$ position}
For the CND$_2$ position we have found the best fits for the water line profiles (red histograms in Fig.~\ref{fig8}) with the derived X$_{\rm H_2O}$ as 1.3$\times$10$^{-6}$ and the derived T$_{\rm d}$ of 15 K for the low--density component, and with the derived X$_{\rm H_2O}$ as 13.3$\times$10$^{-6}$ and the derived T$_{\rm d}$ of 25 K for the high--density component. The H$_2$O line widths observed towards this position are fitted better with the derived V$_{\rm t}$ of 14 km s$^{-1}$.
The modeling of the water lines of this position is complex because towards this line of sight the emission arising from the CND and the 50
km s$^{-1}$ cloud is blended as discussed before.
In Fig. 6 of \citet{Baladron} it can be seen clearly that the CS(1--0) line emission arising from the western edge 
of the 50 km s$^{-1}$ cloud that reaches velocities up to $\sim$80 km s$^{-1}$ is detected in the northeastern CND, region that coincides with our CND$_2$ position.
An H$_2$O abundance of 2$\times$10$^{-10}$ accounts for the emission/absorption contribution from the 50 km s$^{-1}$ cloud in this CND position. We have simulated this cloud having spherical symmetry, a n$_{\rm H_2}$ of 3$\times$10$^5$ cm$^{-3}$, a T$_{\rm k}$ of 100 K and a low T$_{\rm d}$ of 20 K \citep{Baladron,Rod2001,Rod}. The simulated 50 km s$^{-1}$ cloud creates an absorption feature in the modeled 1$_{11}-0_{00}$ and 2$_{12}-0_{01}$ H$_2$O lines (see Fig.~\ref{fig8}). There is also a small difference between the modeled and observed para 1$_{11}-0_{00}$ H$_2$O line intensity at $\sim$100 km s$^{-1}$ (see Fig.~\ref{fig8}). This difference could be caused by either standing waves or the simplicity of spherical symmetry in our models.

Following the same procedure as for the CND$_1$ position, based on $\tau_{\rm d}$(250 $\mu$m) we have estimated a N$_{\rm H_2}$ of 3.2--3.6$\times$10$^{22}$ cm$^{-2}$ for the CND$_2$ position. We have also determined N$_{\rm H_2O}$ of 4.1$\times$10$^{16}$ cm$^{-2}$ for the low--density component and of 4.8$\times$10$^{17}$ cm$^{-2}$ for the high--density component. The whole set of the derived parameters obtained using our model are listed in Table \ref{table2}. The derived N$_{\rm H_2}$ yield a total gas mass of 253 M$_{\odot}$ that is comparable to the mass obtained by \cite{Reque12} for this position.

\subsubsection{20 km s$^{-1}$ cloud position}
As already mentioned we expect that the H$_2$O emission towards the 20 km s$^{-1}$ cloud position is not affected by contributions from other positive--velocity line of sight sources. We have searched for the best fits to the water line profiles varying only the X$_{\rm H_2O}$ and V$_{\rm t}$, finding the values of the X$_{\rm H_2O}$ of 4.0$\times$10$^{-8}$ and V$_{\rm t}$ of 9 km s$^{-1}$.
 
Fig.~\ref{fig8} shows that all the modeled water line profiles in emission/absorption are in 
agreement with the observed line profiles, except in the case of the 1$_{10}-1_{01}$ H$_2^{18}$O spectrum, where the water line is blended with the emission from the $^{13}$CH$_3$OH(1$_{62}-1_{61}$) line. This argument is supported by the 1$_{10}-1_{01}$ H$_2$O to H$_2^{18}$O line intensity ratio that is equal to $\sim$50 at 110 km s$^{-1}$ for the CND$_1$ position, but as low as $\sim$14 for the 20 km s$^{-1}$ cloud position.
The $\tau_{\rm d}$(250 $\mu$m) of 0.08 predicted for this position corresponds to the derived N$_{\rm H_2}$ of 2.7$\times$10$^{23}$ cm$^{-2}$. For this position we have derived a N$_{\rm H_2O}$ as 1.1$\times$10$^{16}$ cm$^{-2}$ included in Table \ref{table2}, where the other derived parameters are also summarized.

\section{Discussion}\label{discu5}
The derived free parameters together with the fixed parameters that provide the best fits to the water line profiles are given in \mbox{Table \ref{table2}}. The derived X$_{\rm H_2O}$, V$_{\rm t}$ and T$_{\rm d}$ are dependent of the assumed source sizes, which are not very well known. However, the V$_{\rm t}$ of 9--23 km s$^{-1}$ obtained for the three studied positions are consistent with those of 15--30 km s$^{-1}$ derived towards the GC \citep{Gusten2}. Furthermore, for the CND we have derived T$_{\rm d}$ of 15--45 K, which agree quite well with the two dust components of 24 and 45 K reported by \citet{Etxaluze}. The T$_{\rm d}$=15$^{+13}_{-13}$ K derived for the low--excitation component in the CND$_2$ position is lower than that derived in the CND$_1$ position. The derived T$_{\rm d}$ of 45$^{+6}_{-3}$ K in the CND$_1$ position is responsible for pumping the 2$_{02}$--1$_{11}$ H$_2$O line (see Sec. \ref{Excitation}), which has an intensity a factor 4 higher than that in the CND$_2$ position.

The derived X$_{\rm H_2O}$ within (0.1--1.3)$\times$10$^{-5}$ for the CND and the derived X$_{\rm H_2O}$ of 4.0$\times$10$^{-8}$ for the 20 km s$^{-1}$ cloud are also consistent with the lower limit of 2$\times$10$^{-8}$ for X$_{\rm H_2O}$ as calculated by \citet{Karlsson} for these two GC sources 
using the non--LTE radiative transfer code RADEX.

The inferred N$_{\rm H_2}$$\sim$(3.0--3.9)$\times$10$^{22}$ cm$^{-2}$ for the CND are similar to those determined from CO in
previous studies \citep{Reque12,Bradford} and also consistent with N$_{\rm H_2}$$\sim$10$^{22}$--10$^{23}$ cm$^{-2}$ calculated from HCN
measurements \citep{Gusten1,Jackson}. The derived N$_{\rm H_2}$ of 2.7$\times$10$^{23}$ for the 20 km s$^{-1}$ cloud is slightly higher than 
that of $\sim$7$\times$10$^{22}$ cm$^{-2}$ derived by \citet{Rod2001} using $^{13}$CO. On the other hand, our derived value of N$_{\rm H_2}$ is lower than
that of $\sim$7$\times$10$^{23}$ cm$^{-2}$ estimated from the ground--state transition of H$^{13}$CO$^+$ for this cloud \citep{Tsuboi}.

\subsection{Excitation of Water}\label{Excitation}
For the CND$_1$ position, we have found that water excitation is affected by the dust emission, since when we have removed the dust effects in our model, the 
observed ortho 1$_{10}-1_{01}$ H$_2$O and H$_2^{18}$O line intensities are slightly overestimated,
the observed para 2$_{02}-1_{11}$ H$_2$O line intensity is underestimated by a factor of $\sim$2 due to the lack of radiative pumping, while the other modeled water lines remain unchanged. 
The 2$_{02}-1_{11}$ H$_2$O line is also found to be pumped through absorption of continuum photons in extragalactic sources \citep{Omont}.
For the CND$_2$ position, there is no strong radiative excitation from the dust emission of the para 2$_{02}-1_{11}$ H$_2$O line as all predicted water line intensities remain unaffected.
Therefore, only collisional excitation is responsible for the 2$_{02}-1_{11}$ H$_2$O line strength.

For the 20 km s$^{-1}$ cloud position, we have noted that all four H$_2$O and the two H$_2^{18}$O lines are also affected by radiative excitation from dust, since all predicted water 
line profiles changed significantly when dust effects were removed in the modeling, with the observed ortho 1$_{10}-1_{01}$ and para 
2$_{02}-1_{11}$ H$_2$O line intensities being overestimated and underestimated by 
a factor of $\sim$2, respectively, due to the lack of dust effects. Therefore, in the CND$_1$ and the 20 km s$^{-1}$ cloud positions the water excitation is determined by collisional effects and absorption of far--infrared continuum photons, while radiation is not important in the water excitation of the CND$_2$ position.

\subsection{Chemistry and heating}
We have derived a X$_{\rm H_2O}$ of 1.3$^{+0.9}_{-0.1}$$\times$10$^{-5}$ for the high--density component of the CND$_2$ position, value that is ten times higher than that derived for the low--density component of the same position. For both density components of the CND$_1$ position we have inferred X$_{\rm H_2O}$ within (6.7--9.3)$\times$10$^{-6}$.

As already mentioned, \citet{Baladron} found that the shock tracer SiO revealed high abundances in the CND, where in contrast the HNCO showed lowest abundances due to its photodissociation by UV photons. 
The spatial correlation between the water and SiO(2--1) emission in the CND (see Sec. \ref{Map557} and \ref{Map988}) points towards grain sputtering as an important mechanism for gas phase water production in the CND. 
We have derived T$_{\rm k}$/T$_{\rm d}$ ratios of 4--17 for the CND, which also supports the idea that mechanical energy from shocks plays a role in the H$_2$O chemistry.

\cite{Harada} studied the chemical composition of the southwest lobe of the CND (our CND$_1$ position) through chemical modeling. Their model considers high temperature chemistry and mimics grain sputtering by shocks and the effects of cosmic--rays. They also studied the effects of UV photons in the chemistry, finding that the abundances of many molecules are not affected in A$_{\rm V}$<1 regions while the H$_3$O$^+$ (that can form H$_2$O via its dissociative recombination \citep{Vejby}) and HCO$^+$ abundances can reach values up to 10$^{-8}$ for A$_{\rm V}$<1 regions. The n$_{\rm H_2}$ and T$_{\rm k}$ of the low--density CND components agree with those modeled by \cite{Harada} for gas with a preshock density of 2$\times$10$^4$ cm$^{-3}$, a shock velocity of 10 km s$^{-1}$ and timescales around 10$^{2.8}$ years after the shock (hereafter scenario 1), while the n$_{\rm H_2}$ and T$_{\rm k}$ of the high--density CND components is in agreement with those modeled for a shocked medium with a preshock density of 2$\times$10$^5$ cm$^{-3}$, shock velocities of 10--20 km s$^{-1}$ and timescales around 10$^{1.2-1.4}$ years (hereafter scenario 2).

The X$_{\rm H_2O}$ of (0.7--1.3)$\times$10$^{-5}$ derived for the high--density components of the CND is a factor 15--29 lower than that of about 2$\times$10$^{-4}$ predicted in scenario 2. Varying cosmic--ray ionization rates within 10$^{-17}$--10$^{-13}$ s$^{-1}$ does not change the X$_{\rm H_2O}$ predicted in scenario 2 (see Fig. 5 of \cite{Harada}).
The above difference can be decreased (within a factor 9--22) when the errors in our estimates are considered. Our X$_{\rm H_2O}$ of (0.7--1.3)$\times$10$^{-5}$ are consistent with those predicted in scenario 2 but with timescales of 10$^4$ years and a cosmic--ray ionization rate of 10$^{-14}$ s$^{-1}$ and assuming that there might have been multiple shocks. This is also in good agreement with the n$_{\rm H_2}$ and T$_{\rm k}$ of the high--density CND components predicted by the models in \cite{Harada}.
The value of 10$^{-14}$ s$^{-1}$ for the cosmic--ionization rate is also consistent with those derived for sources located within the Central Cluster \citep{Goto}.

On the other hand, the X$_{\rm H_2O}$ of 9.3$\times$10$^{-6}$ estimated for the low--density component of the CND$_1$ position is only a factor 2 lower than that predicted in scenario 1 with a preshock density of 2$\times$10$^5$ cm$^{-3}$ and a cosmic--ray ionization rate of 10$^{-16}$ s$^{-1}$ (hereafter modified scenario 1). The X$_{\rm H_2O}$ of 1.3$\times$10$^{-6}$ derived for the low--density component of the CND$_2$ component is 15 times lower than that predicted in modified scenario 1. A cosmic--ionization rate of 10$^{-14}$ s$^{-1}$ decreases the water abundance at timescales around 10$^3$ years, giving a better agreement with our derived water abundances for both low--density CND components in the modified scenario 1.

It is considered that around 14\% and 18\% of the high-- and low--density material, respectively, of the CND could be considered as a PDR (A$_{\rm V}$<5) given their H$_2$ densities and source sizes. The far--ultraviolet radiation field is G$_0$ $\sim$10$^5$ in the inner edge of the CND \citep{Burton}. Using a chemical model, \citet{Hollenbac} derived 
X$_{\rm H_2O}$ of $\sim$10$^{-7}$ for molecular clouds affected by a far--ultraviolet flux of G$_0$<500. But, for G$_0$>500 and the gas density of 10$^4$ cm$^{-3}$ this model predicted
a peak X$_{\rm H_2O}$ around 10$^{-6}$ only at A$_{\rm V}$=8 due to thermal desorption of O atoms and subsequent water production through neutral--neutral reactions, while at A$_{\rm V}$<5 the water is photodissociated and its abundance decreases below 10$^{-8}$. In PDRs the N$_{\rm H_2O}$ are $\sim$10$^{15}$ cm$^{-2}$ \citep{Hollenbac}, which are lower than those derived in the CND at least by a factor of $\sim$41 (see Table \ref{table2}).
The only effect of increasing G$_0$ in the modeling is that the H$_2$O shell penetrates further into the cloud, while the H$_2$O column densities remain
constant \citep{Hollenbac}. From this comparison, PDRs do not seem to play a role in the water chemistry in the CND.

In addition, CO/H$_2$O abundance ratios have been used to establish if there is any PDR contribution to the water emission. The starburst galaxy M82 revealed 
CO/H$_2$O$\sim$40 \citep{Weiss}. The CO lines are a factor of $\gtrsim$50 stronger than the H$_2$O lines in the prototypical galactic PDR Orion 
Bar \citep{Habart}, which is in contrast to what is observed in Mrk 231, where the H$_2$O and CO lines are comparable \citep{Gonzalez10}.
Based on the integrated intensities of $^{13}$CO lines with J=2--1, 6--5, 13-12 obtained by \citet{Reque12}, we have derived CO to H$_2$O(2$_{10}-1_{11}$) integrated line intensity ratios of 4--43 for the CND. The $^{13}$CO data have a similar angular resolution to our H$_2$O(2$_{10}-1_{11}$) data and the $^{13}$CO can be converted to $^{12}$CO by assuming
a $^{12}$C/$^{13}$C=20 ratio, typical for the CG \citep{Wilson}. Our highest CO/H$_2$O ratio of 43 similar to that of M82, could indicate that there is some PDR contribution in the water chemistry of the CND. This result is in contrast to the suggestion found in the previous discussion. A hot CO component found towards the central cavity is heated by a combination of UV photons and shocks \citep{Goicoe2013}.

Apparently the T$_{\rm k}$$<$200 K of the low--density CND components are not high enough for water production through neutral--neutral reactions, which activate at T$_{\rm k}$$>$300 K \citep{Neufeld95}, however, high-temperature chemistry of water could be produced at the shock fronts of the low--density CND components with warmer gas ($>$300 K) and of course in the high--density CND components with T$_{\rm k}$$>$325 K. As mentioned, the high temperature chemistry is considered in the modeling by \cite{Harada}. On the other hand, the T$_{\rm d}$$<$45 K derived in the CND rules out thermal evaporation of H$_2$O because this mechanism needs grain temperatures around 100 K \citep{Fraser}.

It is thought that the effects of X-rays in the CND chemistry are negligible as the X--ray ionization rate is lower than 10$^{-16}$ s$^{-1}$ at H$_2$ column densities > 10$^{21}$ cm$^{-2}$ \citep{Harada}. This is consistent with the results of \cite{Goicoe2013}, who found that X--rays do not dominate the heating of hot molecular gas near Sgr A*.

The X$_{\rm H_2O}$ of $\sim$4.0$\times$10$^{-8}$ derived in the 20 km s$^{-1}$ cloud is at least a factor of $\sim$33 smaller than those derived in the CND, suggesting that the water freeze--out can partially account for the low X$_{\rm H_2O}$. In these regions the water could be produced through an ion--neutral chemistry \citep{Vejby}. It would be interesting to consider if the modeling proposed by \cite{Harada} including cosmic--ray chemistry without shocks would predict the low water abundance derived in the 20 km s$^{-1}$ cloud.

In summary, the X$_{\rm H_2O}$ within (0.1--1.3)$\times$10$^{-5}$ derived in the CND are better explained in scenarios that consider grain sputtering by shocks of 10--20 km s$^{-1}$, cosmic-rays and high temperature chemistry, with a possible contribution of PDR chemistry, while the water freeze--out seems to be responsible for the low X$_{\rm H_2O}$ of 4.0$\times$10$^{-8}$ derived for the 20 km s$^{-1}$ cloud. 

\section{Conclusions}\label{Conclusions}
\begin{enumerate}
\item We presented velocity--integrated ortho 1$_{10}-1_{01}$, and para 2$_{02}-1_{11}$ and 1$_{11}-0_{00}$ water maps of an area 
of $\sim$8$\times$8 pc$^2$ around Sgr A$^{*}$ observed with the {\it Herschel} Space Telescope. The velocity--integrated
ortho 1$_{10}-1_{01}$ and para 1$_{11}-0_{00}$ H$_2$O maps reveal emission/absorption, whereas the para 2$_{02}-1_{11}$ H$_2$O maps reveal
only emission. The ground state ortho water maps show emission in the velocity range of [-95,130] km s$^{-1}$ associated with the CND, the Western Streamer, and the 20 and 50 km s$^{-1}$ clouds. This water emission from the southwest CND and the Western Streamer South is substantially absorbed by foreground sources. The ground state ortho water maps show absorption structures in the velocity range of [-220,10] km s$^{-1}$ associated with foreground sources.
\item The para 2$_{02}$-1$_{11}$ water emission is concentrated towards the CND, the Western Streamer, and the 20 and 50 km s$^{-1}$ clouds.
Based on the lack of absorption of the para 2$_{02}$-1$_{11}$ water emission around the black hole Sgr A$^*$ we
used this emission in a Velocity versus Position Angle diagram, finding that the para 2$_{02}$-1$_{11}$ water emission is tracing the CND and the clouds that are interacting with the SNR Sgr A East.
\item Using a non--local radiative transfer code, we derived X$_{\rm H_2O}$ of $\sim$(0.1--1.3)$\times$10$^{-5}$, V$_{\rm t}$ of 14--23 km s$^{-1}$ and T$_{\rm d}$ of 15--45 K for the CND, and the X$_{\rm H_2O}$ of 4.0$\times$10$^{-8}$ and the V$_{\rm t}$ of 9 km s$^{-1}$ for the 20 km s$^{-1}$ cloud.
From this study, we also found that collisions and dust effects can account for the observed water excitation in the CND$_1$ and the 20 km s$^{-1}$ cloud positions, but there is not need for radiative excitation in the CND$_2$ position.
\item We propose that the gas phase water vapor production in the CND is produces by a combination of grain sputtering by shocks of 10--20 km s$^{-1}$, high temperature and cosmic--ray chemistries plus a probable PDR chemistry, whereas the low X$_{\rm H_2O}$ derived in the 20 km s$^{-1}$ cloud could be a consequence of the water freeze-out.
\end{enumerate}

\begin{acknowledgements}
J.M.--P. and E.G--A. acknowledge partial support by the MINECO and FEDER funding under grants ESP2015--65597--C4--1 and \mbox{ESP2017--86582--C4--1--R}. A. H. acknowledges support for this work by NASA through an award issued by JPL/Caltech. E.G--A. is a Research Associate at the Harvard--Smithsonian Center for Astrophysics. We thank the anonymous referee for the useful comments that improved the manuscript.
\end{acknowledgements}

\newpage

\begin{appendix}
\section{Standing waves in spectra}\label{Standingwave}
\begin{figure}[ht]
\includegraphics[scale=1.8]{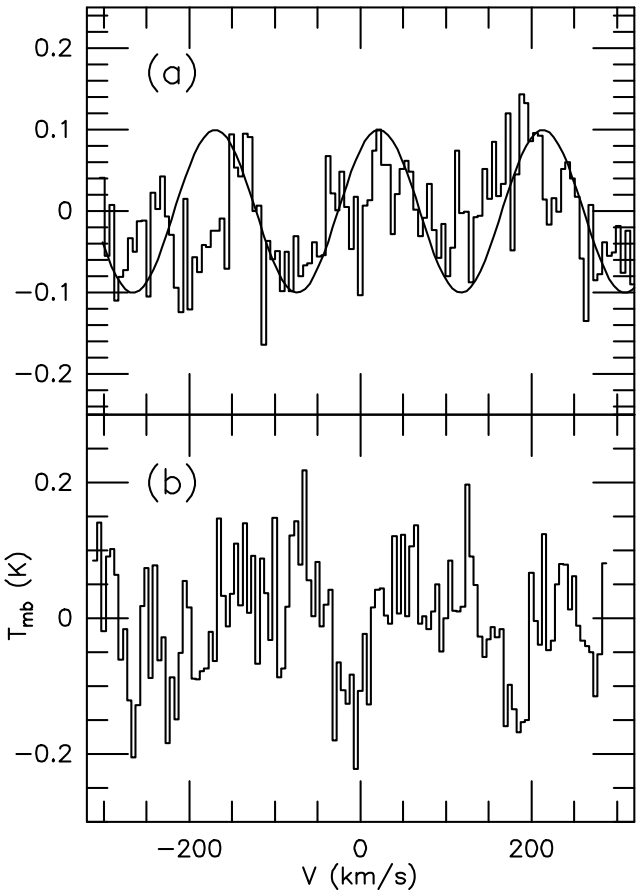}
\caption{Panels (a) and (b) show average spectra extracted from the parallelograms 1 and 2, respectively, shown in Fig.~\ref{fig3}. The spectra are affected by standing waves, with a period around 200 km s$^{-1}$, which originate from the Local Oscillator feed horns of HIFI (see Sec. \ref{Results1}). Panel (a) illustrates a sine function, highlighting the shape of the standing wave.}
\label{standing_wave}
\end{figure}

\section{Physical parameters of the studied sources}\label{Conditions}
\subsection{Dust temperature and turbulent velocity}
\citet{Pierce} derived a fairly uniform dust temperature of 21$\pm$2 K over an extended region in the GC. \citet{Rod} derived T$_{\rm d}$$<$30 K towards molecular clouds distributed along the central 500 pc region of the Galaxy. Additionally, \citet{Etxaluze} using PACS and SPIRE data found that two different components are required to explain the observed continuum in the CND, a hot component with the T$_{\rm d}$ of 45 K and n$_{\rm H_2}$ of 1$\times$10$^4$ cm$^{-3}$ and a cold component with the T$_{\rm d}$ of 24 K and n$_{\rm H_2}$ of 2$\times$10$^{4}$ cm$^{-3}$.
They also found a very hot \mbox{T$_{\rm d}$=90 K} that arises from the central cavity that was not covered by our CND observations. Therefore, this hot T$_{\rm d}$ was not considered in our modeling.  
V$_{\rm t}$ of 15--30 km s$^{-1}$ are estimated towards GC clouds \citep{Gusten2}. Thus, in our models we will explore values of T$_{\rm d}$ of $\sim$15--50 K and V$_{\rm t}$ of 15--30 km s$^{-1}$.

\subsection{H$_2$ density, kinetic temperature and source size}\label{source_size}
\citet{Bradford} studied the CO 
excitation of one CND position (with offsets of $\Delta \alpha=-15\arcsec$ and $\Delta\delta=-30\arcsec$ from Sgr A$^*$), which is close to our studied CND$_1$ position. They used several CO rotational transitions in combination with a LVG radiative transfer code to derive a T$_{\rm k}$ of 240 K and a n$_{\rm H_2}$ of 7.1$\times$10$^4$ cm$^{-3}$. The T$_{\rm k}$ and n$_{\rm H_2}$ derived by \citet{Bradford} are slightly higher than those derived by \citet{Reque12} for the low--excitation gas of the CND. The n$_{\rm H_2}$ around 3$\times$10$^5$ cm$^{-3}$ derived for our two CND positions by \citet{Baladron} agree well with the high--excitation component of \citet{Reque12}, who derived source radii around 0.32 pc and 0.07 pc for the low-- and high--excitation components, respectively. Our two CND positions observed with HIFI coincide with the CND positions studied by \citet{Reque12}. The CND$_1$ and CND$_2$ positions are observed towards the southwest and northeast lobes, respectively, of the CND (see Fig.~\ref{fig6}). As seen in Fig.~\ref{fig6}, the interferometric emission map of CN(2--1) shows that one clump is found towards each CND position. The clumps in the CND$_1$ and CND$_2$ positions are identified as clumps Q and A, respectively, in the interferometric emission map of HCN(4-3) obtained by \cite{Montero09}. Since CO usually traces the H$_2$ column density, we will assume clump radius for the water emission for the CND$_1$ and CND$_2$ positions following the results obtained by \citet{Reque12} (see Table \ref{table2}).
 
The velocity range of [10,40] km s$^{-1}$ of the ortho 1$_{10}-1_{01}$ H$_2$O map in Fig.~\ref{fig4} shows the northern part of the 20 km s$^{-1}$ cloud. For this core we have derived a size around 9 pc for the minor axis, which is somewhat larger than the 7.5 pc size derived by \citet{Zylka}, but the 2$_{02}-1_{11}$ H$_2$O maps show that the 20 and 50 km s$^{-1}$ clouds (see Fig.~\ref{fig4}) have source sizes a factor of $\sim$2 more compact than those observed at 557 GHz. Therefore, for the modeling we will assume a 20 km s$^{-1}$ cloud with spherical symmetry and a size of 4.5 pc. Interferometric maps of ammonia
observed by \citet{Coil} reveal a size for the 20 km s$^{-1}$ cloud relatively similar to that shown by the 2$_{02}-1_{11}$ H$_2$O emission.
\end{appendix}
\end{document}